\documentclass[fleqn]{2017SCGE}
\usepackage{amssymb,amsmath,amsfonts,amsbsy,rotating,bbold}
\usepackage{color,microtype}
\usepackage{graphicx}
\usepackage{tikz}
\usepackage{makecell}
\usepackage{hyperref}

\begin{document}
\ensubject{subject}
\ArticleType{Article}
\SpecialTopic{SPECIAL TOPIC: }
\Year{20??}
\Month{January}
\Vol{60}
\No{1}
\DOI{??????????}
\ArtNo{000000}
\ReceiveDate{January 11, 20??}
\AcceptDate{April 6, 20??}

\title{Constraining the non-Einsteinian polarizations of gravitational waves by pulsar timing array}{Constraining the non-Einsteinian polarizations of gravitational waves by pulsar timing array}

\author[1]{Rui Niu}{nrui@mail.ustc.edu.cn}
\author[2]{Wen Zhao}{wzhao7@ustc.edu.cn}

\AuthorMark{}

\AuthorCitation{}

\address[1]{CAS Key Laboratory for Researches in Galaxies and Cosmology, Department of Astronomy, University of Science and Technology of China, \\ Chinese Academy of Sciences, Hefei, Anhui 230026, China}
\address[2]{School of Astronomy and Space Science, University of Science and Technology of China, Hefei 230026, China}


\abstract{Pulsar timing array (PTA) provides an excellent opportunity to detect the gravitational waves (GWs) in nanoHertz frequency band. In particular, due to the larger number of ``arms" in PTA, it can be used to test gravity by probing the non-Einsteinian polarization modes of GWs, including two spin-1 shear modes labeled by ``$sn$" and ``$se$", the spin-0 transverse mode labeled  by ``$b$" and the longitudinal mode labeled by ``$l$". In this paper, we investigate the capabilities of the current and potential future PTAs, which are quantified by the constraints on the amplitudes parameters $(c_b,c_{sn},c_{se},c_{l})$, by observing an individual supermassive black hole binary in Virgo cluster. We find that for binary with chirp mass $M_{c}=8.77 \times 10^{8} \mathrm{M_{\odot}}$ and GW frequency $f=10^{-9}\mathrm{Hz}$, the PTA at current level can detect these GW modes if $c_b > 0.00106$, $c_l > 0.00217$, $c_{se} > 0.00271$, $c_{sn} > 0.00141$, which will be improved by about two orders if considering the potential PTA in SKA era. Interesting enough, due to effects of the geometrical factors, we find that in SKA era, the constraints on the $l$, $sn$, $se$ modes of GWs are purely dominated by several pulsars, instead of the full pulsars in PTA.}



\maketitle


\begin{multicols}{2}

\section{Introduction}
Einstein's general theory of relativity (GR) is thought as the most successful theory to describe gravitational physics. Since GR was proposed, it has been strictly tested in various cases, and the predictions of GR have received experimental confirmation in high precision \cite{will2006confrontation} \cite{will2018theory}. However, the difficulty of quantization, as well as the problems in galactic dynamics and cosmology, motivate people to try to find feasible alternative gravity theories. It is well known that there exist two polarizations of gravitational waves $h_+$ and $h_{\times}$ in GR. However, in metric theories of gravity, gravitational waves (GWs) can have up to six possible polarizations. Massless scalar-tensor gravitational waves contain the transverse breathing mode \cite{zhang2017testing}. In massive scalar-tensor theories, the longitudinal mode exist \cite{liu2018waveforms}. More general metric theories predict additional longitudinal modes, up to the full complement of six \cite{will2006confrontation} \cite{will2018theory}. Thus, GR and its alternatives can be tested by measuring the polarization properties of GWs.

Since the detection of GW event GW150914 in 2015 \cite{abbott2016observation}, LIGO and Virgo collaborations have directly detected six GWs events \cite{abbott2016gw151226, scientific2017gw170104, abbott2017gw170814, abbott2017gw170817, abbott2017gw170608, li2017, gao2018}. The direct detections of GWs inaugurate the new era of gravitational-wave astronomy. However, the two detectors of LIGO have been constructed to have their respective arms as parallel as possible. While this configuration maximizes the joint sensitivity of the two detectors to gravitational waves, their ability to detect different modes of polarization is minimized. The pulsar timing array (PTA), which is constituted by an array of millisecond pulsars (MSP) observed in a long time, offers an alternative way to detect GWs in the nanohertz band (1-100nHz) \cite{sazhin1978opportunities, detweiler1979pulsar}. Unlike interferometric detectors, a PTA includes a large number of ``arms''. Although there will be a network of interferometric detectors in the near future, the directional complexity makes PTA superior for detecting different polarization modes of GWs.
	
	In previous work \cite{lee2008pulsar}, the sensitivity of PTA to the other four possible non-Einsteinian polarizations of isotropic stochastic GWs background has been considered. In this paper, we focus on the continuous GWs from individual sources. We analyze the capabilities of current and future PTAs to constrain the non-Einsteinian polarizations. The sources considered in this paper are inspiralling supermassive black hole binaries (SMHBs) in circular orbits. The frequency evolution over a typical PTA data span is smaller than the frequency resolution of PTA observation. In addition, binary systems spend most of their lifetime in slowly evolving period. Monochromatic binaries can represent the majority of detectable sources. Therefore, we neglect the frequency evolution of binaries in the analyses.
	
	The outline of this paper is as follows. In section 2, we review the response of the timing residuals obtained in pulsar timing experiments to single-source GWs considered in this paper. In particular, we generalize this result to all the non-Einsteinian polarizations of GWs. In section 3, we analyze the capabilities of current and future PTAs to constrain the non-Einsteinian polarizations. The summary and discussion of the results is given in section 4.

\section{Pulsar timing residuals caused by gravitational waves}
	The radio pulses from pulsars, especially millisecond pulsars, have an extremely stable regularity. The null geodesics of radio waves from pulsars can be perturbed by GWs, which induce the fluctuations of pulse arrival times. The timing residuals, which are obtained by subtracting out all known effects from observed pulse arrival times, contain the information of GWs. If the measurements of timing residuals are precise enough, it is possible to directly detect the GWs. In this section, we will present the derivation of Detweiler's formulae from which we can obtain the relationship between the metric perturbation and the timing residuals measured in pulsar timing experiments. Meanwhile, we extend these results to all the other non-Einsteinian polarizations of GWs.

\subsection{GWs in GR}	
	Let us first consider two Einsteinian modes of GWs. The following results have derived in  \cite{detweiler1979pulsar, anholm2009optimal}. In this paper, we summary these calculations and extend them to the general polarization modes of GWs. In linearized theory with the transverse and traceless gauge, we have the metric
	\begin{equation}
	g_{\mu \nu}=\eta_{\mu \nu} + h_{\mu \nu}
	\ , \qquad
	g^{\mu \nu}=\eta^{\mu \nu} - h^{\mu \nu} ,
	\end{equation}
	\begin{equation}
	h_{\mu \nu} =
		\begin{pmatrix}
		0&0&0&0\\0&h_+&h_\times&0\\0&h_\times&-h_+&0\\0&0&0&0
		\end{pmatrix}.
	\end{equation}
In Minkowksi space-time, the null vector of radio waves from pulsar to earth can be written as
	\begin{equation}
	s^\mu = \nu (1, -\alpha, -\beta, -\gamma),
	\end{equation}
where $\alpha$, $\beta$, $\gamma$ are the direction cosines with the $\hat{x}$, $\hat{y}$, $\hat{z}$ directions respectively, and $\nu$ is the frequency of pulse.
In perturbed space-time, photon's null vector becomes $\sigma^\mu$ which satisfy
	\begin{equation}
	g_{\mu \nu} \sigma^\mu \sigma^\nu= 0 .
	\end{equation}
In the linear theory, this equation can be written as,
	\begin{equation}
	(\eta_{\mu \nu} + h_{\mu \nu})(s^\mu + \delta s^\mu)(s^\nu + \delta s^\nu) = 0.
	\end{equation}
Ignoring the higher order terms, we obtain
	\begin{equation}
	\delta s^\alpha = -\frac{1}{2} \eta^{\alpha \mu} h_{\mu \nu} s^\nu.
	\end{equation}
Thus, the null vector in perturbed space-time becomes
	\begin{equation}
	\sigma^\mu = s^\mu - \frac{1}{2} \eta^{\mu \alpha} h_{\alpha \beta} s^\beta,
	\end{equation}
which can be explicitly computed in perturbed space-time,
	\begin{equation}\label{sigma^mu}
	\sigma^\mu = \nu \Bigl(1, \quad[(\frac{h_+}{2}-1)\alpha + \frac{h_\times}{2} \beta], \quad[\frac{h_\times}{2}\alpha - (\frac{h_+}{2}+1) \beta], \quad -\gamma \Bigr).
	\end{equation}
The $t$-component of $\sigma^{\mu}$ satisfies geodesic equation as follows,
	\begin{align}
	\frac{\mathrm{d}\sigma^0}{\mathrm{d}\lambda} &= -\frac{1}{2} h_{\alpha \beta, 0} \sigma^\alpha \sigma^\beta \\ &= -\frac{1}{2} (\dot{h}_+\sigma^1\sigma^1 + 2\dot{h}_{\times}\sigma^1\sigma^2 - \dot{h}_+\sigma^2\sigma^2),
	\end{align}
i.e.,
	\begin{equation}\label{dnu/dt}
	\frac{\mathrm{d}\nu}{\mathrm{d}\lambda} = -\frac{1}{2} \bigl[\dot{h}_+\nu^2(\alpha^2-\beta^2) + 2\dot{h}_{\times}\nu^2\alpha\beta \bigr].
	\end{equation}
First, let us consider a special case that gravitational waves propagate along $\hat{z}$ direction, i.e. $h_{\mu \nu}(t,z) = h_{\mu \nu}(t-z)$.
So, we can have
	\begin{equation}\label{dh+/dlamdba}
	\frac{\mathrm{d}h_+}{\mathrm{d}\lambda} = \frac{\partial h_+}{\partial t} \frac{\mathrm{d}t}{\mathrm{d}\lambda} + \frac{\partial h_+}{\partial z} \frac{\mathrm{d}z}{\mathrm{d}\lambda},
	\end{equation}
and
	\begin{equation}
	\frac{\partial h_+}{\partial z} = -\frac{\partial h_+}{\partial t}.
	\end{equation}
Because of $\displaystyle \sigma^\mu = {\mathrm{d}x^{\mu}}/{\mathrm{d}\lambda}$, we obtain that
	\begin{align}
	\label{dt/dlambda}\frac{\mathrm{d}t}{\mathrm{d}\lambda} &= \nu, \\
	\label{dz/dlambda}\frac{\mathrm{d}z}{\mathrm{d}\lambda} &= \frac{\mathrm{d}z}{\mathrm{d}t} \frac{\mathrm{d}t}{\mathrm{d}\lambda} = -\gamma\nu.
	\end{align}
Using relations in (\ref{dt/dlambda}) and (\ref{dz/dlambda}), the equation (\ref{dh+/dlamdba}) becomes
	\begin{equation}
	\frac{\mathrm{d}h_+}{\mathrm{d}\lambda} = \dot{h}_+\nu + \dot{h}_+\gamma\nu.
	\end{equation}
Similarly, we also obtain
	\begin{equation}
	\frac{\mathrm{d}h_\times}{\mathrm{d}\lambda} = \dot{h}_\times\nu + \dot{h}_\times\gamma\nu.
	\end{equation}
The time derivatives in equation (\ref{dnu/dt}) as derivatives with respect to the affine parameter $\lambda$ are given by,
	\begin{equation}
	\begin{aligned}
	\label{dt}
	\dot{h}_+ &= \frac{1}{\nu+\gamma\nu} \frac{\mathrm{d}h_+}{\mathrm{d}\lambda}, \\
	\dot{h}_\times &= \frac{1}{\nu+\gamma\nu} \frac{\mathrm{d}h_\times}{\mathrm{d}\lambda}.
	\end{aligned}
	\end{equation}
Substituting above equations into equation (\ref{dnu/dt}), we get
	\begin{equation}
	\frac{1}{\nu}\frac{\mathrm{d}\nu}{\mathrm{d}\lambda} = -\frac{1}{2} \Bigl[ \frac{(\alpha^2-\beta^2)}{1+\gamma}\frac{\mathrm{d}h_+}{\mathrm{d}\lambda} + 2\frac{\alpha\beta}{1+\gamma}\frac{\mathrm{d}h_\times}{\mathrm{d}\lambda} \Bigr].
	\end{equation}
Integrating from pulsar to earth along the geodesic curve,
we derive the final result,
	\begin{equation}
	\frac{\nu(t)-\nu_0}{\nu_0} = -\frac{1}{2}\frac{(\alpha^2-\beta^2)}{1+\gamma}\Delta h_+ - \frac{\alpha\beta}{1+\gamma}\Delta h_\times,
	\end{equation}
where $\nu(t)$ is the observed frequencies of pulsar, $\nu_0$ is the emitted frequencies, and $\Delta h_{+,\times} = h_{+,\times}^e - h_{+,\times}^p$ is the difference between the metric perturbation at the pulsar and the observer.

	Now we will derive the contribution from gravitational waves in arbitrary direction. It is a generalization of the results derived previously. As before, we consider the $\sigma^{0}$ component of geodesic equation, and only preserve the first order of perturbation,
	\begin{align}
	\frac{\mathrm{d}\sigma^0}{\mathrm{d}\lambda} = -\frac{1}{2} h_{\alpha \beta, 0} s^\alpha s^\beta.
	\end{align}
We can write $s^\mu$ as $s^\mu = \nu(1, -\hat{p})$, where the $\hat{p}$ is the direction vector of the pulsar. So
	\begin{equation}\label{geodesic_eq}
	\frac{\mathrm{d}\sigma^0}{\mathrm{d}\lambda} = -\frac{1}{2}\nu^2 h_{ij, 0}\hat{p}^i\hat{p}^j,
	\end{equation}
where $i$ and $j$ are spatial indices. The metric perturbation of gravitational waves propagating along $\hat{\Omega}$ direction has the form $h_{\mu\nu}(t-\hat{\Omega}\cdot\vec{x})$. We replace the time derivatives in equation (\ref{geodesic_eq}) with the $\lambda$ derivatives as follows,
	\begin{align}
	\label{dt_dlambda}
	\frac{\mathrm{d}h_{\mu\nu}(t-\hat{\Omega}\cdot\vec{x})}{\mathrm{d}\lambda} &= \frac{\partial h_{\mu\nu}(t-\hat{\Omega}\cdot\vec{x})}{\partial t} \frac{\mathrm{d}t}{\mathrm{d}\lambda} + \frac{\partial h_{\mu\nu}(t-\hat{\Omega}\cdot\vec{x})}{\partial \hat{\Omega}\cdot\vec{x}} \frac{\mathrm{d}\hat{\Omega}\cdot\vec{x}}{\mathrm{d}\lambda} \notag \\
	&= \frac{\partial h_{\mu\nu}(t-\hat{\Omega}\cdot\vec{x})}{\partial t} \nu - \frac{\partial h_{\mu\nu}(t-\hat{\Omega}\cdot\vec{x})}{\partial t} \hat{\Omega} \frac{\mathrm{d}\vec{x}}{\mathrm{d}\lambda} \notag \\
	&= \frac{\partial h_{\mu\nu}(t-\hat{\Omega}\cdot\vec{x})}{\partial t} (1+\hat{\Omega}\cdot\hat{p}) \nu.
	\end{align}
It should be noted that, in perturbed space-time, the null vector of photon is the function of the affine parameter instead of constant. The relationship, $s^\mu = \mathrm{d}x^{\mu} / \mathrm{d}\lambda = \nu(1, -\hat{p})$, is valid only in Minkowski space-time. In perturbed space-time, there is a small difference between $\mathrm{d}\vec{x} / \mathrm{d}\lambda$ and $-\hat{p}$. But the deviation is in terms of $\mathcal{O}(h)$, and we have neglected this deviation in the derivation of equation (\ref{dt_dlambda}). Substituting (\ref{dt_dlambda}) into geodesic equation (\ref{geodesic_eq}), we obtain
	\begin{equation}
	\frac{\mathrm{d}\sigma^0}{\mathrm{d}\lambda} = -\frac{1}{2} \nu^2 \frac{1}{(1+\hat{\Omega}\cdot\hat{p}) \nu} \frac{\mathrm{d}h_{ij}(t-\hat{\Omega}\cdot\vec{x})}{\mathrm{d}\lambda} \hat{p}^i\hat{p}^j,
	\end{equation}
	\begin{equation}
	\frac{1}{\nu}\frac{\mathrm{d}\nu}{\mathrm{d}\lambda} = -\frac{1}{2} \Bigl[ \frac{\hat{p}^i\hat{p}^j}{1+\hat{\Omega}\cdot\hat{p}} \frac{\mathrm{d}h_{ij}(t-\hat{\Omega}\cdot\vec{x})}{\mathrm{d}\lambda} \Bigr].
	\end{equation}
Integrating from pulsar to earth as previous, we can get
	\begin{equation}\label{redshift}
	\frac{\nu(t)-\nu_0}{\nu_0} = -\frac{1}{2} \Bigl[ \frac{\hat{p}^i\hat{p}^j}{1+\hat{\Omega}\cdot\hat{p}} \Delta h_{ij}(t-\hat{\Omega}\cdot\vec{x}) \Bigr].
	\end{equation}
In practice, the timing residual $R(t)$ that we can obtain in pulsar timing experiments, is defined as (\ref{redshift}):
	\begin{equation}
	R(t) = \int_0^t \frac{\nu(t')-\nu_0}{\nu_0} \, \mathrm{d}t'.
	\end{equation}

\subsection{GWs in the general metric theory}
In a general metric theory, gravitational waves can have up to six possible polarization modes as follows,
	\begin{equation}
	h_{\mu \nu} =
		\begin{pmatrix}
		0&0&0&0\\
		0&h_b+h_+&h_\times&h_{sn}\\
		0&h_\times&h_b-h_+&h_{se}\\
		0&h_{sn}&h_{se}&h_l
		\end{pmatrix}.
	\end{equation}
The ``$+$'' and ``$\times$'' denote the two Einsteinian polarization modes; the ``$sn$'' and ``$se$'' denote the two spin-1 shear modes; the ``$l$'' and ``$b$'' denote the spin-0 longitudinal mode and the spin-0 breathing mode.
It has been shown that the result (\ref{redshift}) derived above is valid for other four polarization modes as well \cite{lee2008pulsar}. It is convenient to use polarization tensors $\epsilon^{A}_{\mu\nu}(\hat{\Omega})$ (with $A = +, \times, b, se, sn, l$ labeling the polarizations) to describe the metric perturbation,
	\begin{equation}
	h_{\mu\nu} = \sum_{A} \epsilon^{A}_{\mu\nu}(\hat{\Omega}) h^A(t-\hat{\Omega}\cdot\vec{x}).
	\end{equation}	
The equation (\ref{redshift}) can be written as
	\begin{equation}
	\frac{\nu(t)-\nu_0}{\nu_0} = -\frac{1}{2} \sum_A \Bigl[ \frac{\hat{p}^i\hat{p}^j}{1+\hat{\Omega}\cdot\hat{p}} \epsilon^{A}_{ij}(\hat{\Omega}) \cdot \Delta h^A(t-\hat{\Omega}\cdot\vec{x}) \Bigr].
	\end{equation}
For convenience, we define the geometrical term:
	\begin{equation}\label{geometry_term}
	F^A = \frac{\hat{p}^i\hat{p}^j}{1+\hat{\Omega}\cdot\hat{p}} \epsilon^{A}_{ij}(\hat{\Omega}),
	\end{equation}
which depends on the position of the GW source, the Earth and the pulsar, as well as the polarization mode of GW.
	\begin{figure}[H]
	\centering
	\begin{tikzpicture}
	\draw[->, color = blue, thick] (0,0) -- (3,0);
	\node [below] at (3,0) {$y$};
	\draw[->, color = blue, thick] (0,0) -- (0,3);
	\node [left] at (0,3) {$z$};
	\draw[->, color = blue, thick] (0,0) -- (-1.5,-1.5);
	\node [right] at (-1.5,-1.5) {$x$};
	
	\draw[dashed] (0,0) -- (1.723,3.888);
	\fill (1.148,2.592) circle (0.04);
	
	\draw[->] (1.148,2.592) -- (1.4926, 3.3696);
	\node [right] at (1.4926, 3.3696) {$\hat{r}$};
	\draw[->] (1.148,2.592) -- (1.9542, 2.8982);
	\node [below] at (1.9542, 2.8982) {$\hat{\alpha}$};
	\draw[->] (1.148,2.592) -- (0.5511, 3.2451);
	\node [below] at (0.5511, 3.2451) {$\hat{\delta}$};
	
	\draw[dashed] (0,0) -- (-2.25775, 0.86725);
	\draw[->] (0,0) -- (-0.9031,0.3469);
	\node [below] at (-0.9031,0.3469) {$\hat{p}$};
	\end{tikzpicture}
	\caption{The coordinate system: $\hat{p}$ is the direction of a pulsar, $\hat{r}$ is the direction of a gravitational source (note: $\hat{r} = -\hat{\Omega}$), $\hat{\alpha}$ and $\hat{\delta}$ are unit vectors orthogonal to the propagation direction and to each other, $\hat{\alpha}$ points to increasing right ascension, $\hat{\delta}$ points to increasing declination.}
	\label{coordinate system}
	\end{figure}
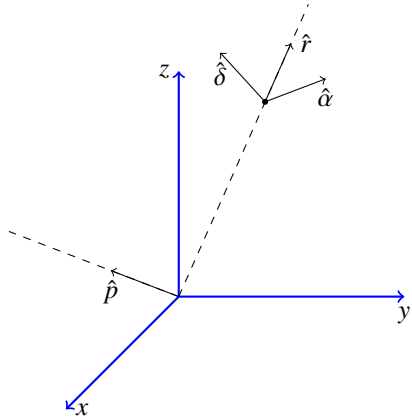

For a given coordinate system, the triad $(\hat{x},\hat{y},\hat{z})$ is the fundamental celestial frame. $(\alpha,\delta)$ and $(\alpha_p,\delta_p)$ are the right ascension and declination of GW source and pulsar, respectively. Then, 	
	\begin{subequations}
	\begin{align}
	\hat{p} &= (\cos{\delta_p}\cos{\alpha_p},\ \cos{\delta_p}\sin{\alpha_p},\ \sin{\delta_p}), \\
	\hat{r} &= (\cos{\delta}\cos{\alpha},\ \cos{\delta}\sin{\alpha},\ \sin{\delta}), \\
	\hat{\alpha} &= (-\sin{\alpha},\ \cos{\alpha},\ 0), \\
	\hat{\delta} &= (-\sin{\delta}\cos{\alpha},\ -\sin{\delta}\sin{\alpha},\ \cos{\delta}).
	\end{align}
	\end{subequations}
Now, we can explicitly write the polarization tensors as follows:
	\begin{equation}
	\begin{aligned}
	\epsilon^{+}_{ij} &= \hat{\alpha}_i\hat{\alpha}_j - \hat{\delta}_i\hat{\delta}_j, \
	& \epsilon^{\times}_{ij} &= \hat{\alpha}_i\hat{\delta}_j - \hat{\alpha}_i\hat{\delta}_j, \
	& \epsilon^{b}_{ij} &= \hat{\alpha}_i\hat{\alpha}_j + \hat{\delta}_i\hat{\delta}_j, \\
	\epsilon^{sn}_{ij} &= \hat{\alpha}_i\hat{r}_j + \hat{r}_i\hat{\alpha}_j, \
	& \epsilon^{se}_{ij} &= \hat{\delta}_i\hat{r}_j + \hat{r}_i\hat{\delta}_j, \
	& \epsilon^{l}_{ij} &= \hat{r}_i\hat{r}_j.
	\end{aligned}
	\end{equation}
It is straightforward to calculate geometry terms (\ref{geometry_term}) of the four non-Einsteinian polarization modes,
	\begin{subequations}\label{geometry_term_s}
		\begin{equation}
		\begin{split}
F^b = &\frac{1}{2(1-\cos \theta)} \biggl\{ \cos^2(\delta_p) \Bigl( -\cos^2(\delta) \cos[2(\alpha-\alpha_p)]+\sin^2(\delta)+1 \Bigr) \\
&+ 2\cos^2(\delta) \sin^2(\delta_p)	- \sin(2\delta)\sin(2\delta_p)\cos(\alpha -\alpha_p) \biggr\},
		\end{split}
		\end{equation}
		\begin{equation}
		\begin{split}
F^{sn} = &\frac{1}{(1-\cos \theta)} \biggl\{ -\sin(\delta)\sin(2\delta_p)\sin(\alpha - \alpha_p) \\
&- \cos(\delta)\cos^2(\delta_p)\sin[2 (\alpha-\alpha_p)] \biggr\},
		\end{split}
		\end{equation}
		\begin{equation}
		\begin{split}
F^{se} = &\frac{1}{(1-\cos \theta)} \biggl\{ \cos(2\delta)\sin(2\delta_p)\cos(\alpha - \alpha_p) \\
&- \frac{1}{4} \sin(2\delta) \Bigl\{\cos (2\delta_p) \Bigl(\cos[2(\alpha-\alpha_p)]+3\Bigr) - 2\sin^2(\alpha-\alpha_p)\Bigr\}\biggr\},
		\end{split}
		\end{equation}
		\begin{equation}
F^{l} = \frac{1}{(1-\cos \theta)} [\cos(\delta )\cos(\delta_p)\cos(\alpha-\alpha_p) + \sin(\delta)\sin(\delta_p)]^2,
		\end{equation}
	\end{subequations}
where $\theta$ is the opening angle between the GW source and pulsar with respect to the observer,
	\begin{equation}
	\cos \theta = \hat{r}\cdot\hat{p} = \cos(\delta)\cos(\delta_p)\cos(\alpha-\alpha_p) + \sin(\delta)\sin(\delta_p).
	\end{equation}
The GW strain term $\Delta h^A$ is the difference between the metric perturbation at the geodesic ending, i.e. the Earth term, $h^A(t_e-\hat{\Omega}\cdot\vec{x}_e)$, and the metric perturbation at the geodesic beginning, i.e. the pulsar term, $h^A(t_p-\hat{\Omega}\cdot\vec{x}_p)$. We consider a frame in which
	\begin{equation}
	\begin{aligned}
	t_e &= t, \qquad & \vec{x}_e &= 0, \\
	t_p &= t - d_p, \qquad & \vec{x}_p &= d_p\hat{p},
	\end{aligned}
	\end{equation}
where $d_p$ is the pulsar distance. The GW strain term is written as:	
	\begin{equation}
	\Delta h^A = h^A(t) - h^A[t-d_p(1-\cos\theta)].
	\end{equation}
The specific functional forms of the strain depend on the type of sources. In this paper, we consider supermassive black hole binaries (SMBHBs) in circular orbits, and neglect the frequency evolution. In the general modified gravity, the waveforms of non-Einsteinian modes have the following forms \cite{hansen2015projected},
	\begin{subequations} \label{GW_strain_nonE}
	\begin{align}
	&h^b = c_b h_0 \sin(\iota) \cos(\Phi_{orb}), \\
	&h^{sn} = c_{sn} h_0 \sin(\Phi_{orb}), \\
	&h^{se} = c_{se} h_0 \cos(\iota) \cos(\Phi_{orb}), \\
	&h^l = -2 c_b h_0 \sin(\iota) \cos(\Phi_{orb}),
	\end{align}
	\end{subequations}
to leading PN order, where $\iota$ is the inclination angle of the binary orbital angular momentum with respect to the line of sight, $\Phi_{orb}$ is the orbital phase. The quantities $c_A$ are parameters of four polarizations, which depends on the theory of gravity. In this paper, for the general metric of gravity, we set them as the free parameters, and are constrained by pulsar timing experiments. And these constraints have been easily translated to the constraint on the parameters of the gravitational theories. $h_0$ defined as
	\begin{equation}
	h_0 = 4 \frac{(GM_c)^{\frac{4}{3}} (2\pi f_{orb})^{\frac{1}{3}}}{r}
	\end{equation}
in which $r$ is the distance to the GWs source, $f_{orb}$ is the orbital frequency, $M_c$ is the binary chirp mass defined as $M_c^{5/3} = m_1m_2(m_1+m_2)^{-1/3}$ with $m_1$ and $m_2$ being the binary component masses. Now we can write down the pulsar timing residuals for four non-Einsteinian polarizations, respectively
	\begin{subequations} \label{timing_residual}
	\begin{align}
	R^b &= c_b \cdot F^b \frac{h_0}{2\pi f} \sin(\iota) \Bigl\{\sin(2\pi f t)-\sin \bigl(2\pi f [t-d_p(1-\cos\theta)]\bigl)\Bigr\}, \\
	R^{sn} &= - c_{sn} \cdot F^{sn} \frac{h_0}{2\pi f} \Bigl\{\cos(2\pi f t)-\cos \bigl(2\pi f [t-d_p(1-\cos\theta)]\bigl)\Bigr\}, \\
	R^{se} &= c_{se} \cdot F^{se} \frac{h_0}{2\pi f} \cos(\iota) \Bigl\{\sin(2\pi f t)-\sin \bigl(2\pi f [t-d_p(1-\cos\theta)]\bigl)\Bigr\}, \\
	R^l &= -2 c_l \cdot F^l \frac{h_0}{2\pi f} \sin(\iota) \Bigl\{\sin(2\pi f t)-\sin \bigl(2\pi f [t-d_p(1-\cos\theta)]\bigl)\Bigr\},
	\end{align}
	\end{subequations}
where we have set the polarization angle and phase constant to zero.

\section{The capabilities of current and future PTAs to constrain the non-Einsteinian polarizations}

In this section, we will analyze the capabilities of current and future PTAs to constrain the parameters ($c_b$, $c_{sn}$, $c_{se}$, $c_l$) of equations (\ref{timing_residual}) in various cases. From equations (\ref{timing_residual}), we can see that timing residuals depend on chirp mass, orbital frequency and inclination angle of the binary, as well as distances and celestial positions of the binary and the pulsars. In this work, we consider the influence of chirp mass, orbital frequency and inclination angle on the capabilities of PTAs to constrain the non-Einsteinian polarizations. For the binary, It has been shown that the Virgo cluster may represent a gravitational waves hotspot for PTAs searches \cite{Simon2014Gravitational}, so we consider a circular binary system with distance $r = 16.5 \ \mathrm{Mpc}$ and celestial position $(\alpha, \delta) = (3.2594 \ \mathrm{rad}, 0.2219 \ \mathrm{rad})$ located in the Virgo cluster as in the previous work \cite{zhu2016detection}. In our analyses, we assume an observation with 10-year span and 2-week cadence, which determines the sensitivity frequency band \cite{perera2018improving}. The lower frequency limit is $1/T_{span}$, where $T_{span}$ is the total observational span. The upper frequency limit depends on the sampling pattern. In our analyses, the observational data are assumed to be evenly sampled. So, we can simply use Nyquist frequency $1/2T_{cadence}$ as the upper limit, where $T_{cadence}$ is the two-week interval between two observations.

The detection threshold is set to $\rho = 10$. $\rho$ is the signal-to-noise ratio defined as \cite{zhu2016detection} :
	\begin{equation}
	\label{SNR}
	\rho^2 = \sum_{j=1}^{N_p}\rho_j^2 = \sum_{j=1}^{N_p}\sum_{i=1}^{N_{obs}} \Bigl[\frac{R^A(t_i)}{\sigma_j}\Bigr]^2,
	\end{equation}
where $R^A(t)$ is the timing residual caused by GWs as shown in equations (\ref{timing_residual}) ($A$ denotes the polarization modes which can be $+, \times, b, se, sn, l$), $\sigma_j$ is the noise RMS of $j$-th pulsar, $N_{obs}$ is the total number of sampling points, $N_p$ is the number of pulsars, $\rho_j$ is the individual signal-to-noise ratio for each pulsar. The capabilities of PTAs to constrain the non-Einsteinian polarizations are represented by the quantities of parameters ($c_b$, $c_{sn}$, $c_{se}$, $c_l$) that can make signal-to-noise ratio just reach the detection threshold.

	\begin{table}[H]
	\centering
	\caption{The RMS timing residual levels and distances for the 30 pulsars in the simulated PTA date sets, which are taken from the previous work \cite{zhu2016detection}. The RMS timing residuals broadly represent the actual measurements of current PTA data sets. The distances of pulsars come from ATNF Pulsar Catalogue \cite{manchester2005ATNF}. Note that distances are poorly known for most of pulsars listed here.}
	\resizebox{\columnwidth}{!}{
	\begin{tabular}{lcc|lcc}
		\Xhline{1.2pt}
		\bfseries Pulsar & \bfseries RMS(ns) & \bfseries $\boldsymbol{d_p}$(kpc) & \bfseries Pulsar & \bfseries RMS(ns) & \bfseries $\boldsymbol{d_p}$(kpc) \\
		\Xhline{0.6pt}
		J0437$-$4715 &  58  & 0.16  & J1600$-$3053  & 202 & 2.40 \\
  		J1640$+$2224 &  158 & 1.19  & J1713$+$0747  & 116 & 1.05 \\
  		J1741$+$1351 &  233 & 0.93  & J1744$-$1134  & 203 & 0.42 \\
  		J1909$-$3744 &  102 & 1.26  & J1939$+$2134  & 104 & 5.00 \\
  		J2017$+$0603 &  238 & 1.32  & J2043$+$1711  & 170 & 1.13 \\
  		J2241$-$5236 &  300 & 0.68  & J2317$+$1439  & 267 & 1.89 \\
  		J0023$+$0923 &  320 & 0.95  & J0030$+$0451  & 723 & 0.28 \\
  		J0613$-$0200 &  592 & 0.90  & J1017$-$7156  & 500 & 0.26 \\
  		J1024$-$0719 &  846 & 0.49  & J1446$-$4701  & 500 & 2.03 \\
  		J1614$-$2230 &  336 & 1.77  & J1738$+$0333  & 316 & 1.47 \\
  		J1832$-$0836 &  577 & 1.40  & J1853$+$1303  & 369 & 1.60 \\
  		J1857$+$0943 &  505 & 0.90  & J1911$+$1347  & 500 & 1.60 \\
  		J1918$-$0642 &  547 & 1.40  & J1923$+$2515  & 535 & 0.99 \\
  		J2010$-$1323 &  733 & 1.29  & J2129$-$5721  & 880 & 0.40 \\
  		J2145$-$0750 &  535 & 0.57  & J2214$+$3000  & 399 & 1.32 \\
		\Xhline{1.22pt}
	\end{tabular}
	}
	\label{table_IPTA-pulsar}
	\end{table}

\subsection{Constraints by present PTAs}
We use the same simulated PTA data sets in the previous work \cite{zhu2016detection} to represent current PTAs. The PTA date sets consist of 30 pulsars as listed in table (\ref{table_IPTA-pulsar}). The RMS timing residuals broadly represent the actual measurements of current PTA data sets. The distances and celestial positions of pulsars come from ATNF Pulsar Catalogue\footnote{\url{http://www.atnf.csiro.au/research/pulsar/psrcat}} \cite{manchester2005ATNF}.

	\emph{The influence of inclination $\iota$.}
We first consider the influence of inclination angle on the capabilities of current PTAs. For the chirp mass and the frequency, as in the previous work \cite{zhu2016detection}, we consider the strong signal with $M_{c}=8.77 \times 10^{8} \mathrm{M_{\odot}}$ and the weak signal with $M_{c}=1.93 \times 10^{8} \mathrm{M_{\odot}}$, and two typical cases $f=10^{-8}\mathrm{Hz}$ and $f=10^{-9}\mathrm{Hz}$, i.e. in the following calculations, we discuss the following four sets of signals,
	\begin{enumerate}
	\item $M_{c}=1.93 \times 10^{8} \mathrm{M_{\odot}}$, $f=10^{-8}\mathrm{Hz}$;
	\item $M_{c}=1.93 \times 10^{8} \mathrm{M_{\odot}}$, $f=10^{-9}\mathrm{Hz}$;
	\item $M_{c}=8.77 \times 10^{8} \mathrm{M_{\odot}}$, $f=10^{-8}\mathrm{Hz}$;
	\item $M_{c}=8.77 \times 10^{8} \mathrm{M_{\odot}}$, $f=10^{-9}\mathrm{Hz}$.
	\end{enumerate}
	
\begin{figure*}[h]
	\centering
	\includegraphics[scale=0.5]{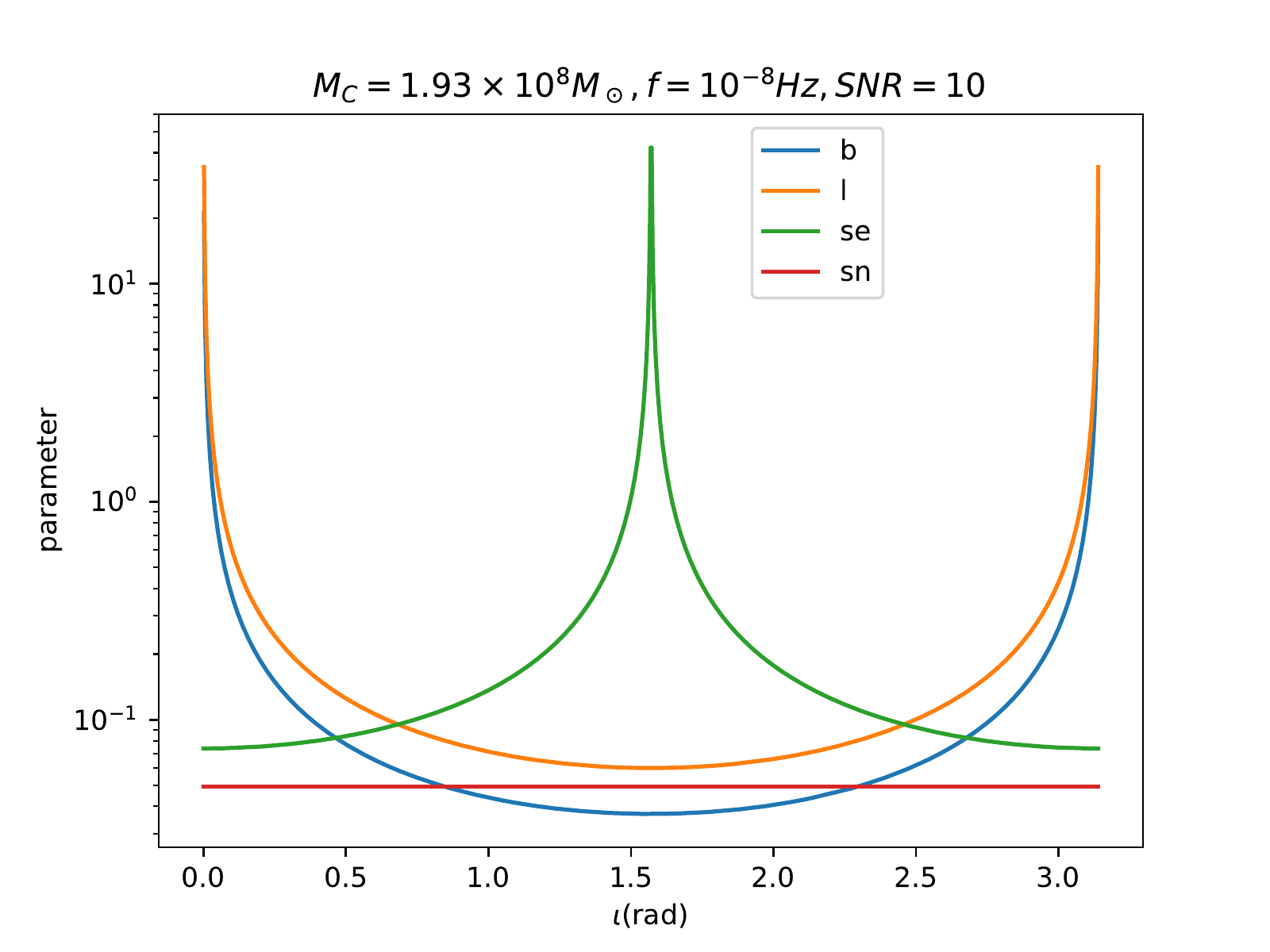}
	\includegraphics[scale=0.5]{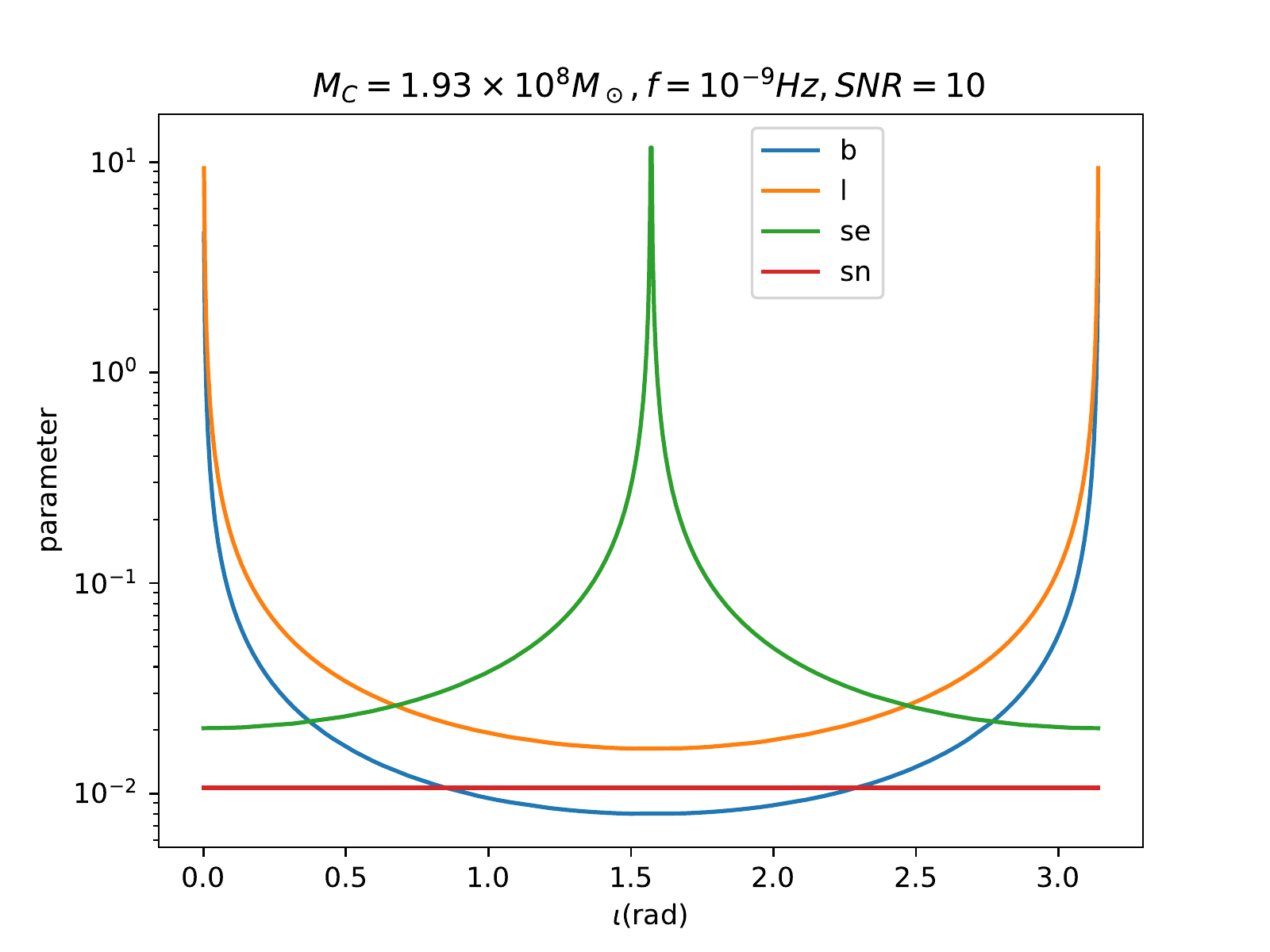} \\
	\includegraphics[scale=0.5]{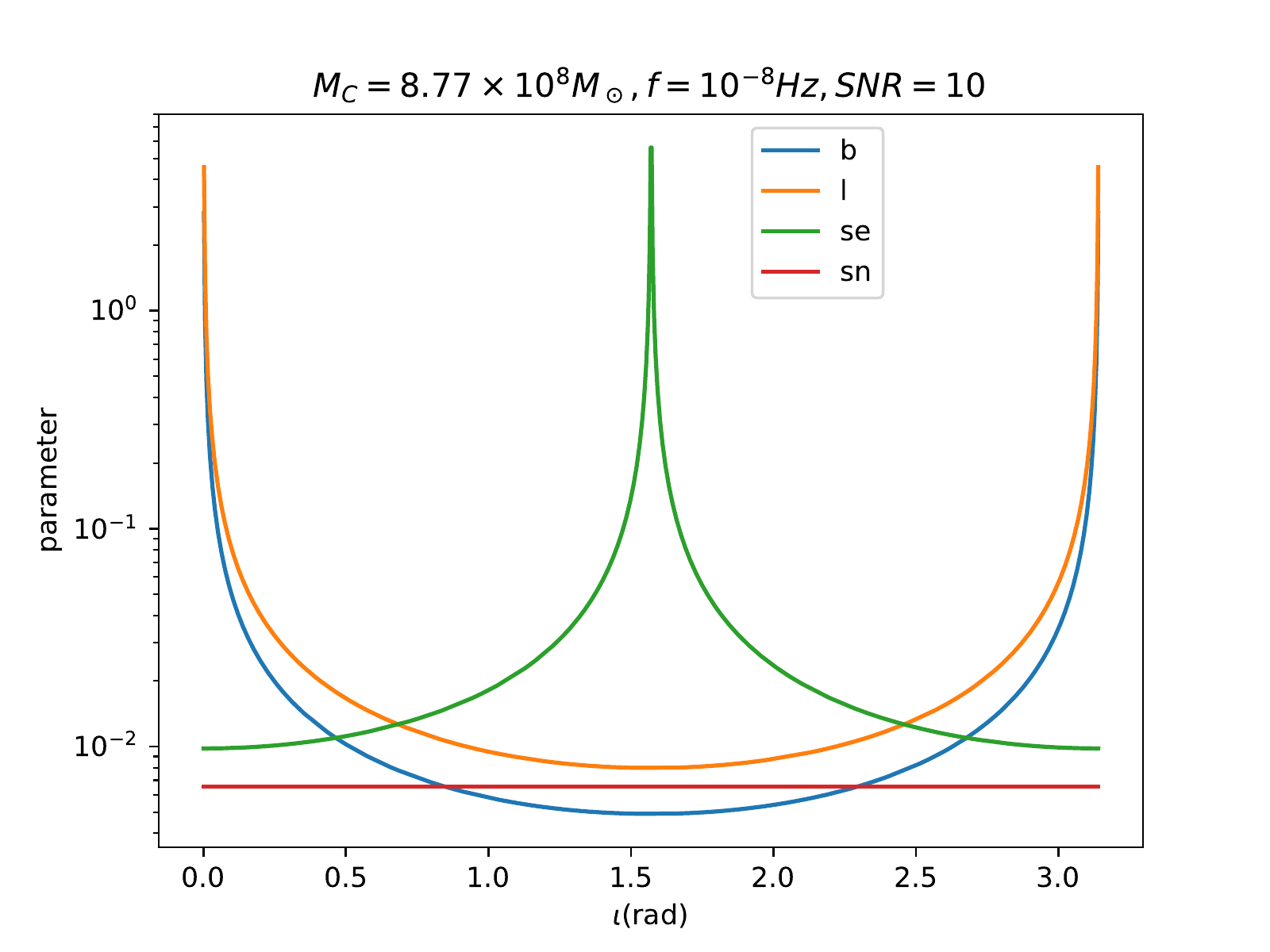}
	\includegraphics[scale=0.5]{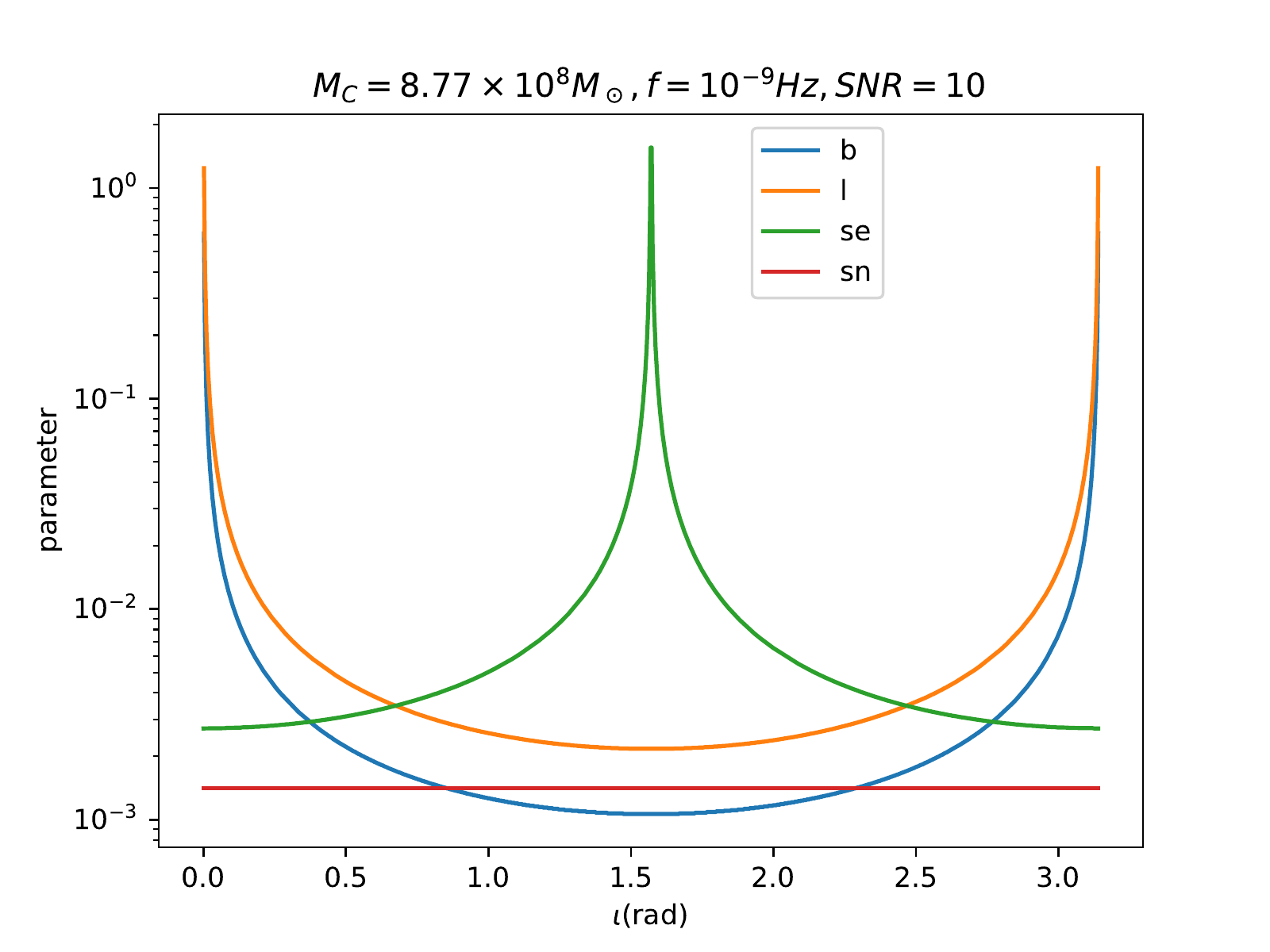}
	\caption{The influence of inclination angle on the capabilities of current PTAs to constrain the non-Einsteinian polarizations. The horizontal axes denote the inclination angle $\iota$ which is defined as the angle of the binary orbital angular momentum with respect to the line of sight. The vertical axes denote the constraints of the parameters ($c_b$, $c_{sn}$, $c_{se}$, $c_l$) in equations (\ref{timing_residual}) which represent the capabilities of PTAs to constrain the non-Einsteinian polarizations. The constraints of the parameters are determined by making signal-to-noise ratio (\ref{SNR}) just reach the detection threshold which is set to $\rho = 10$. The colors denote the different polarizations. As in the previous work \cite{zhu2016detection}, we consider a circular binary system located in the Virgo cluster with distance $r = 16.5 \ \mathrm{Mpc}$ and celestial position $(\alpha, \delta) = (3.2594 \ \mathrm{rad}, 0.2219 \ \mathrm{rad})$ as the GWs source. We use the same simulated PTA data sets taken from the previous work \cite{zhu2016detection} to represent the current PTAs. Four sets of signals are considered: (\romannumeral1). $M_{c}=1.93 \times 10^{8} \mathrm{M_{\odot}}$, $f=10^{-8}\mathrm{Hz}$; (\romannumeral2). $M_{c}=1.93 \times 10^{8} \mathrm{M_{\odot}}$, $f=10^{-9}\mathrm{Hz}$; (\romannumeral3). $M_{c}=8.77 \times 10^{8} \mathrm{M_{\odot}}$, $f=10^{-8}\mathrm{Hz}$; (\romannumeral4). $M_{c}=8.77 \times 10^{8} \mathrm{M_{\odot}}$, $f=10^{-9}\mathrm{Hz}$.}
	\label{ipta_i}
	\end{figure*}

The results are shown in figure (\ref{ipta_i}). The horizontal axes denote the inclination angle $\iota$ which is defined as the angle of the binary orbital angular momentum with respect to the line of sight. The vertical axes denote the constraints of the parameters ($c_b$, $c_{sn}$, $c_{se}$, $c_l$) in equations (\ref{timing_residual}) which represent the capabilities of PTAs to constrain the non-Einsteinian polarizations. The colors denote the different polarizations. Different polarizations have different dependence on inclination. The constraints of $b$, $l$ modes are best when the inclination angle is $90^\circ$ and are divergent when the inclination angle is $0^\circ$ or $180^\circ$. The constraints of $se$ mode are best when the inclination angle is $0^\circ$ or $180^\circ$ and are divergent when the inclination angle is $90^\circ$. The $sn$ mode has no relations to the inclination angle. These characters are coincident with the equation (\ref{timing_residual}). The timing residuals of $b$, $l$ modes depend on inclination angle by sine function, the timing residuals of $se$ mode depend on inclination angle by cosine function and the timing residuals of $se$ mode are independent of inclination angle.

With the optimal inclination angle, for case 1, the constraints of four non-Einsteinian polarizations are: $c_b = 0.0369$, $c_l = 0.0599$, $c_{se} = 0.0737$, $c_{sn} = 0.0492$; for case 2, the constraints are: $c_b = 0.00799$, $c_l = 0.0163$, $c_{se} = 0.0204$, $c_{sn} = 0.0106$; for case 3, the constraints are: $c_b = 0.00490$, $c_l = 0.00796$, $c_{se} = 0.00979$, $c_{sn} = 0.00654$, and for case 4, the constraints become: $c_b = 0.00106$, $c_l = 0.00217$, $c_{se} = 0.00271$, $c_{sn} = 0.00141$. By the comparison between the upper panels and the low panels, the constraints are better for lower frequency. By the comparison between the left panels and the right panels, the constraints are better for the more massive binary. These are consistent with the prediction of equations (\ref{timing_residual}), which show that the timing residuals depend on chirp mass by $M_c^{\frac{4}{3}}$, and depend on frequency by $f^{-\frac{2}{3}}$.

	\emph{The influence of frequence $f$.}
The sensitivity frequency band is determined by the span and cadence of the observation \cite{perera2018improving}. We assume an observation with 10-year span and 2-week cadence, which corresponds to the frequency band $3.17 - 414\mathrm{nHz}$. We consider the chirp mass $M_{c}=1.93 \times 10^{8} \mathrm{M_{\odot}}$ which was used in the previous work \cite{zhu2016detection}. The distance and celestial position of the binary, as well as the PTA data sets remain the same with the case that we have considered above. Differently, here we consider the cases with various $f$.

\begin{figure*}[hp]
	\centering
	\includegraphics[scale=0.5]{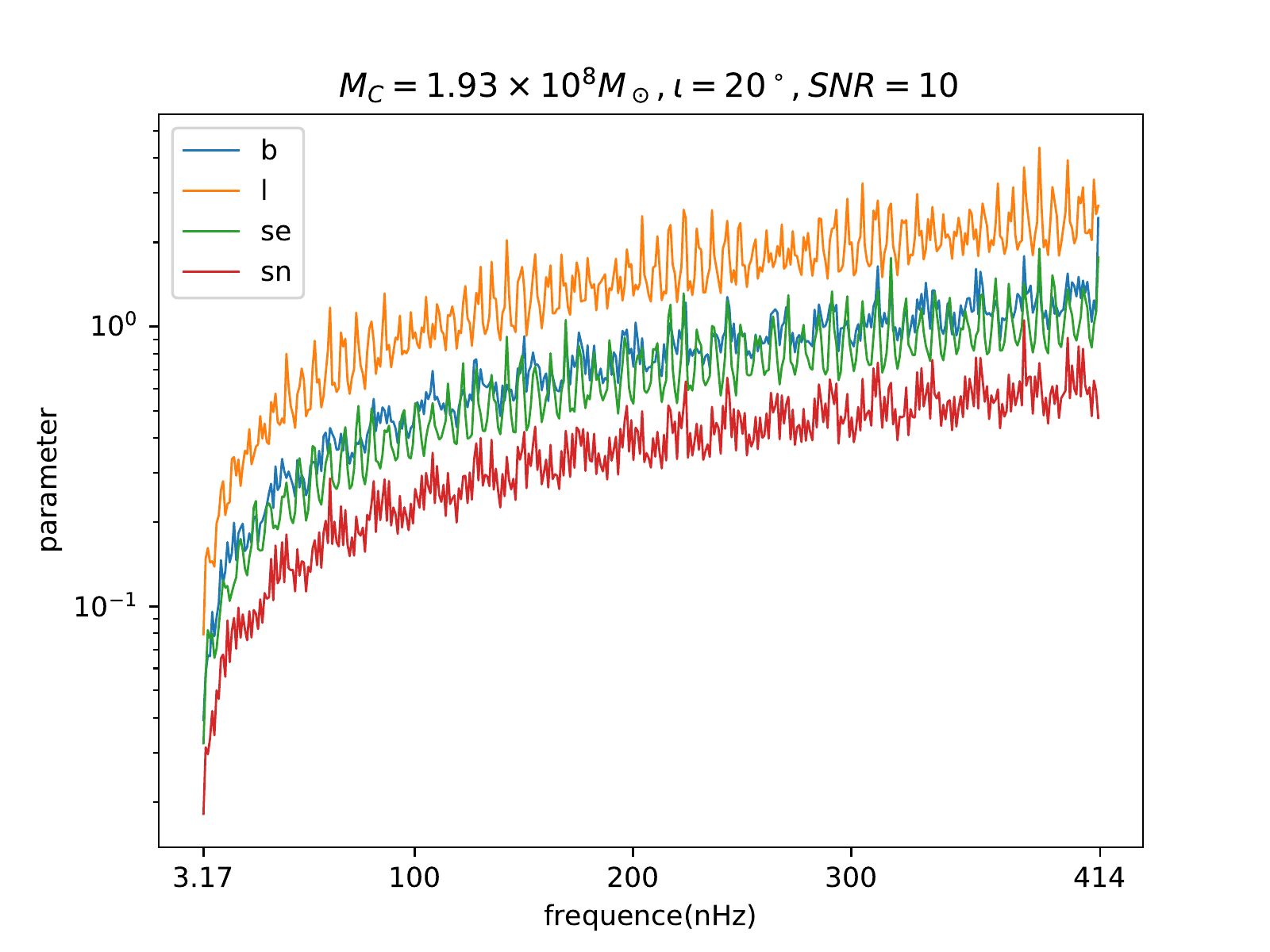}
	\includegraphics[scale=0.5]{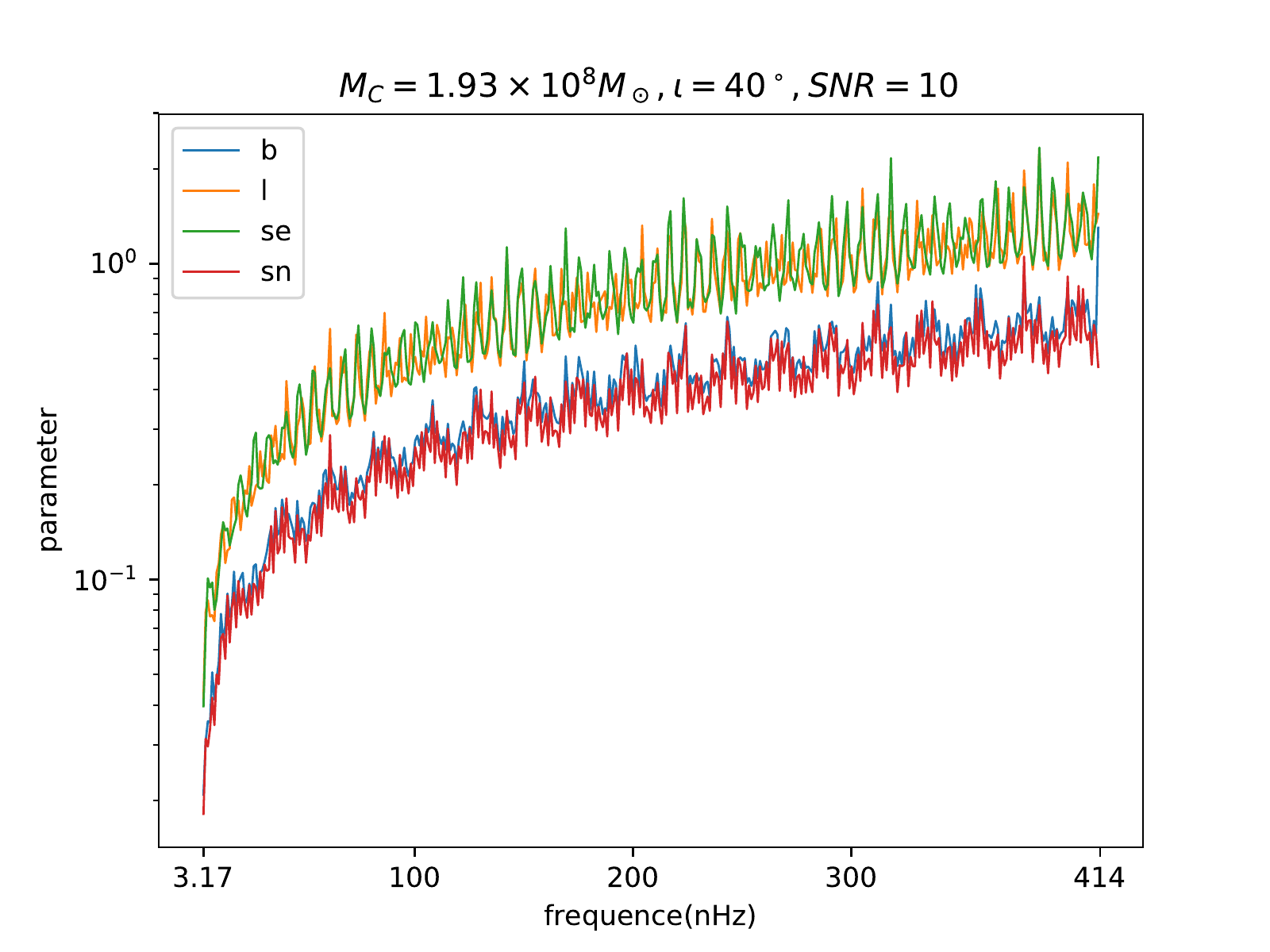}
	\includegraphics[scale=0.5]{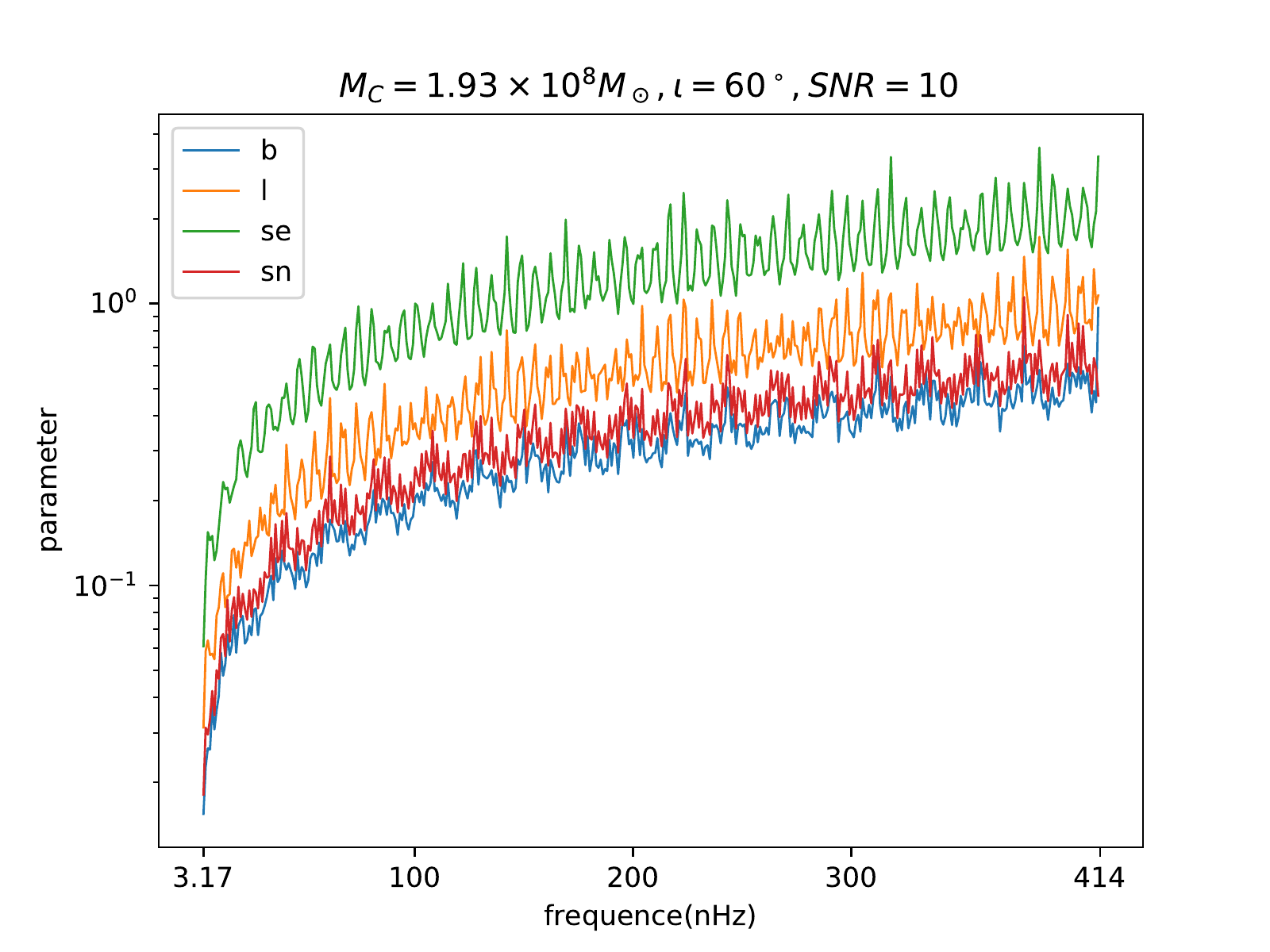}
	\includegraphics[scale=0.5]{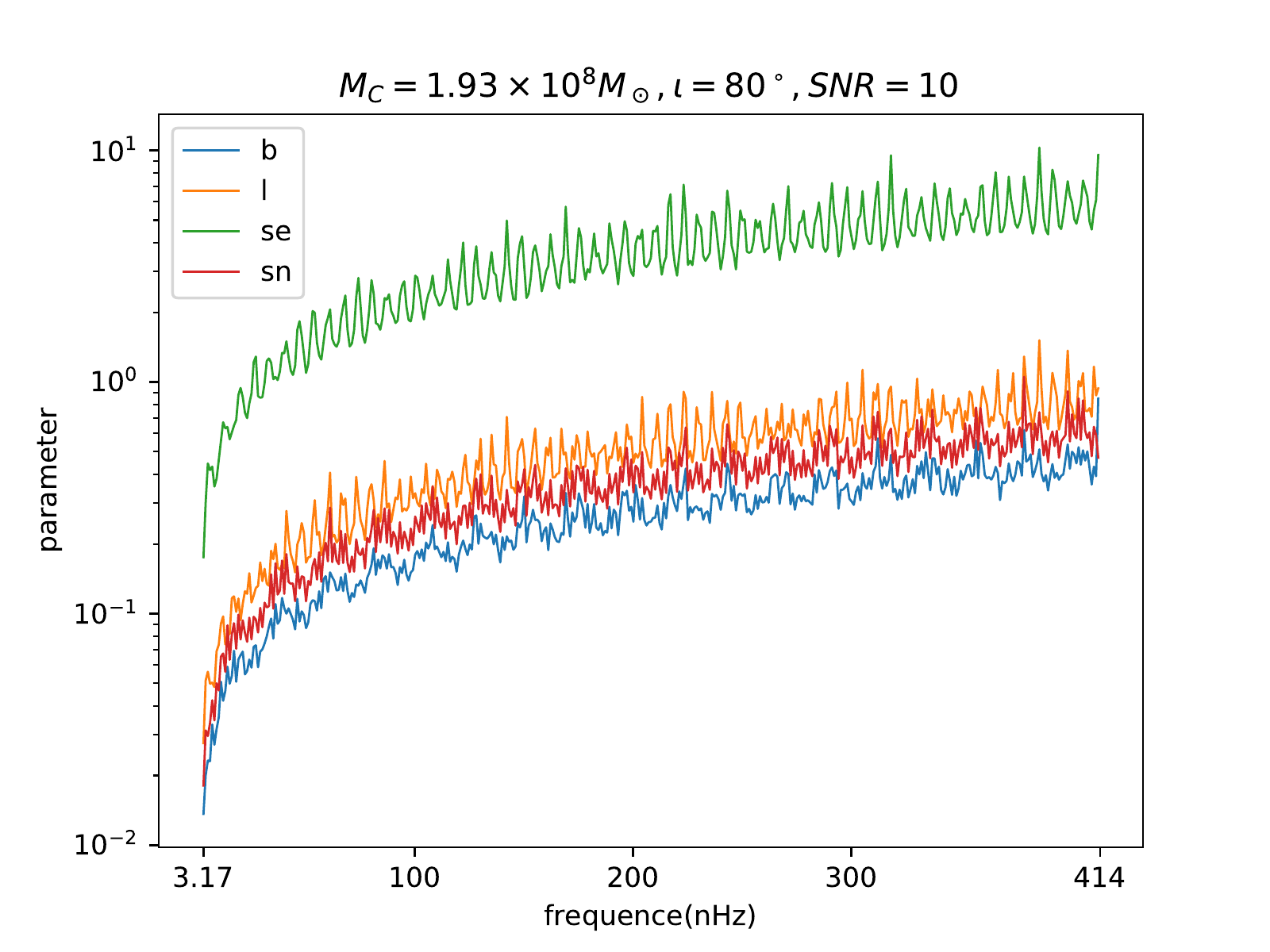}
	\newpage
	\caption{The influence of frequency on the capabilities of current PTAs to constrain the non-Einsteinian polarizations. The horizontal axes denote the frequency of the GWs which is also the orbital frequency for the non-Einsteinian polarizations (\ref{GW_strain_nonE}). The sensitivity frequency band is determined by the pattern of the observation. In our analyses, we assume an observation with 10-year span and 2-week cadence which correspond to the frequency band $3.17 - 414\mathrm{nHz}$. The vertical axes denote the constraints of the parameters ($c_b$, $c_{sn}$, $c_{se}$, $c_l$) in equations (\ref{timing_residual}) which represent the capabilities of PTAs to constrain the non-Einsteinian polarizations. The constraints of the parameters are determined by making signal-to-noise ratio (\ref{SNR}) just reach the detection threshold. The distances and celestial positions of the GWs source and the pulsars, as well as the detection threshold remain the same with the previous case at which we have considered the influence of inclination angle. }
	\label{ipta_f}
	\end{figure*}

The results are shown in figure (\ref{ipta_f}). It is noticeable that the constraints have some fluctuations when we consider the influence of frequency. We can see that from the equation (\ref{timing_residual}), the frequency is different from the chirp mass and the inclination angle, it appears not only in the amplitude term but also in the phase term. The signal-to-noise ratio (\ref{SNR}) is the accumulation of the contributions of every pulsar in the PTA. When we increase the frequency, the parts of amplitude terms in equation (\ref{timing_residual}) vary monotonously, but the parts in the braces of the equation (\ref{timing_residual}) contribute irregular changes to the signal-to-noise ratio. As a result, we can see jagged lines in the figure (\ref{ipta_f}). Since the timing residuals (\ref{timing_residual}) depend on the frequency by $f^{-\frac{2}{3}}$, it is natural that the constraints are better for lower frequency. In the figures, as the inclination angle increasing, the lines of $sn$ mode keep stationary, the lines of $b$ and $l$ modes move down, the lines of $se$ mode move upwards. This feature is coincident with the figure (\ref{ipta_i}).

\begin{figure*}[hp]
	\centering
	\includegraphics[scale=0.5]{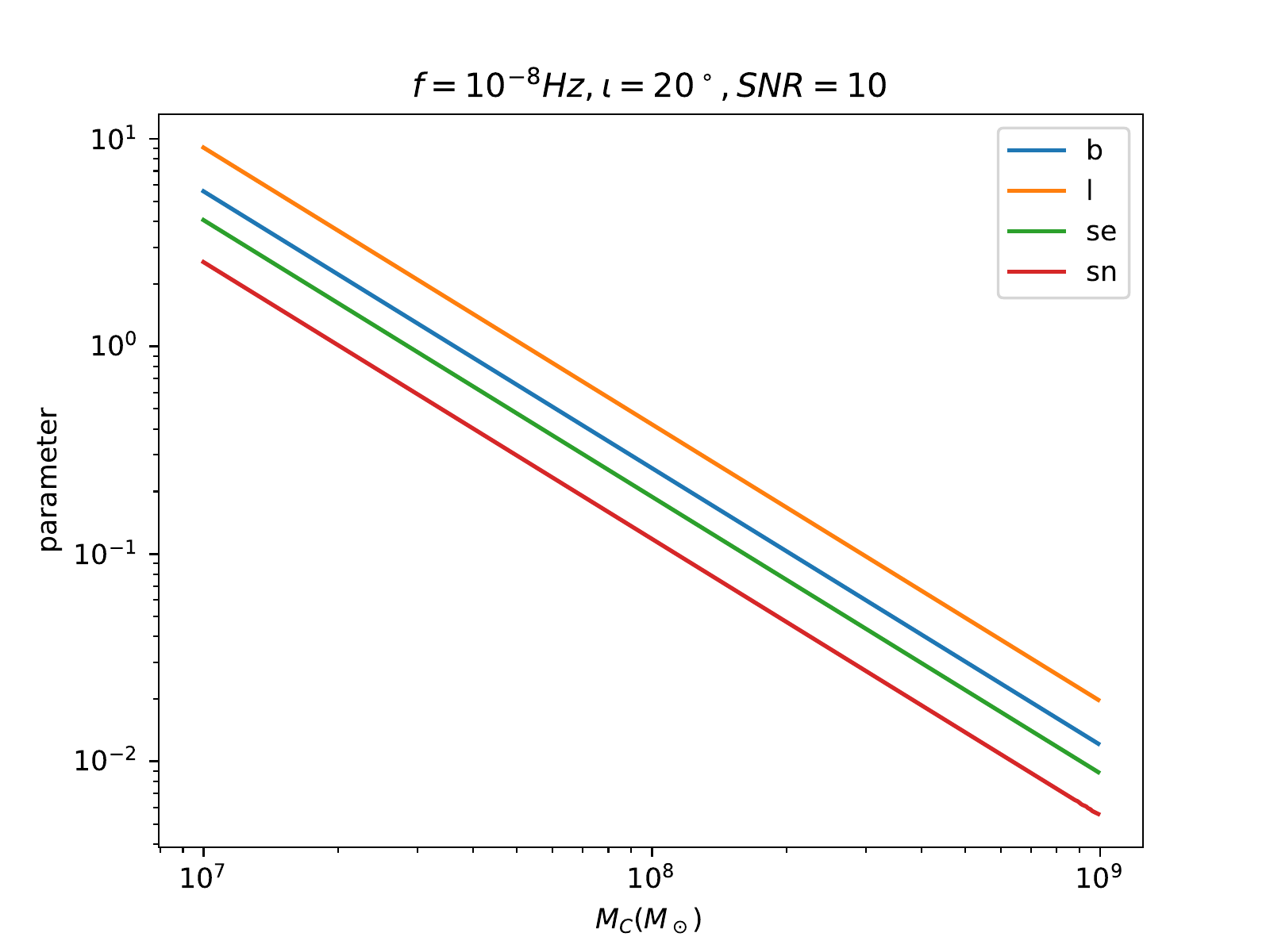}
	\includegraphics[scale=0.5]{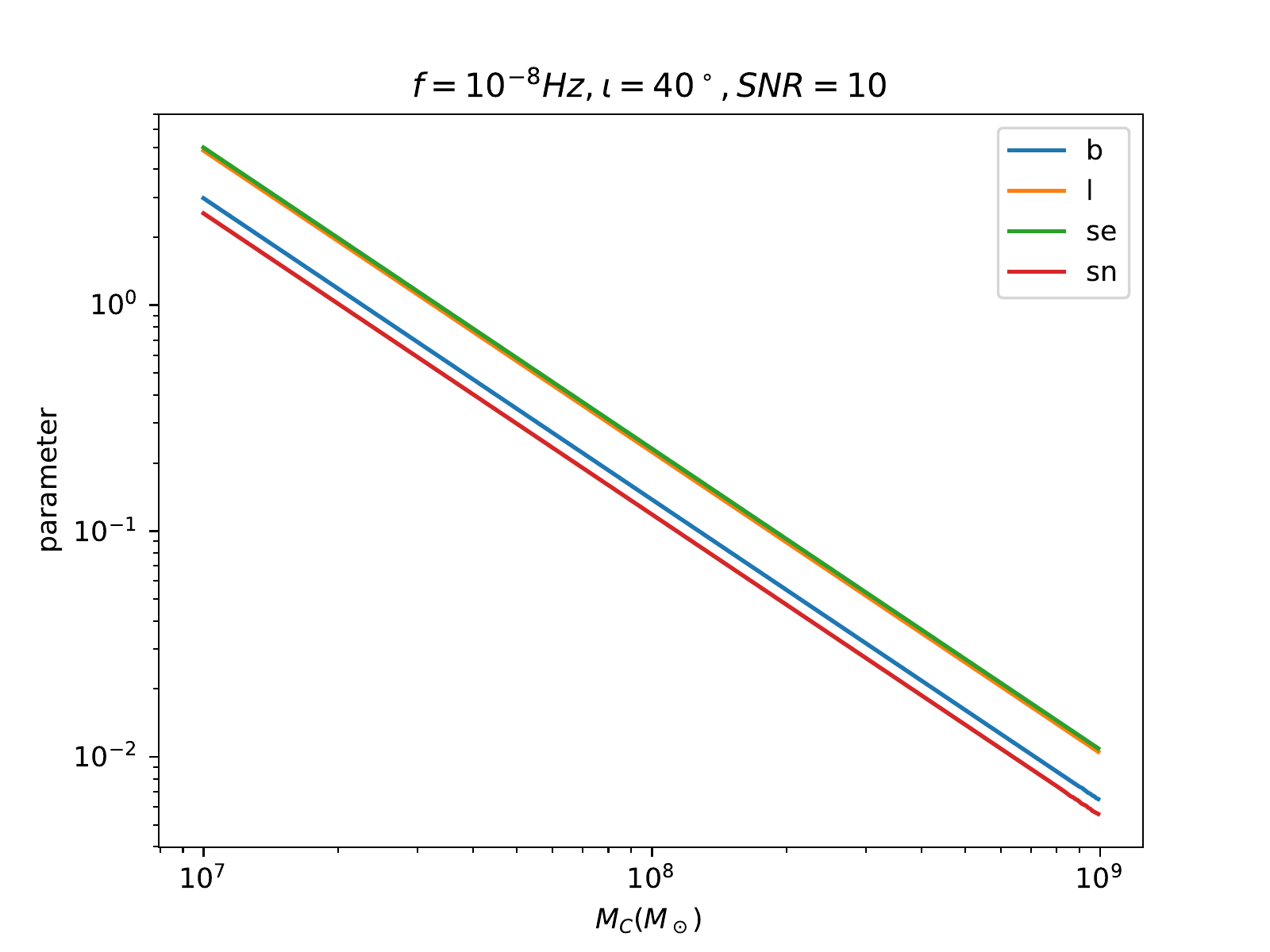}
	\includegraphics[scale=0.5]{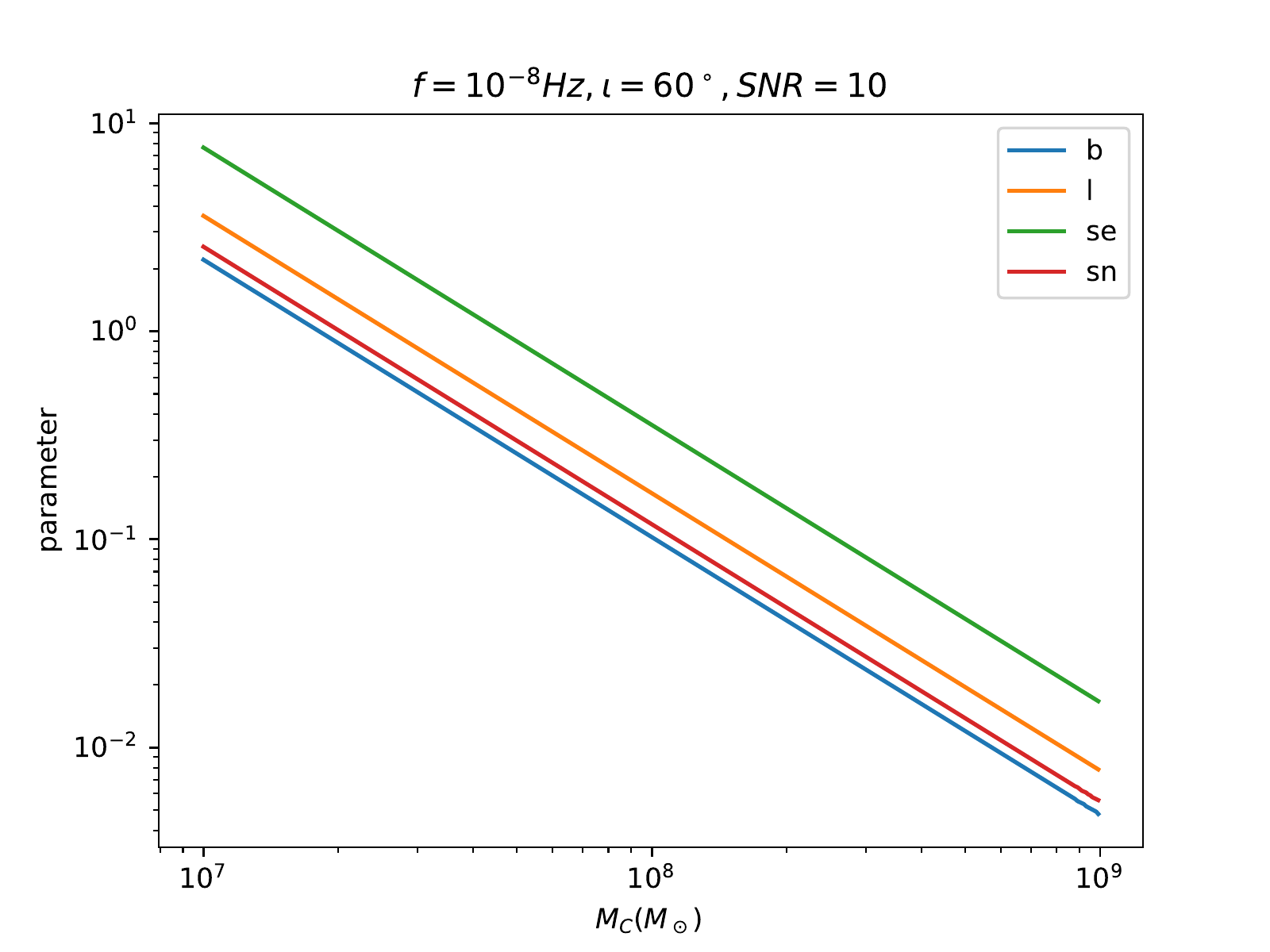}
	\includegraphics[scale=0.5]{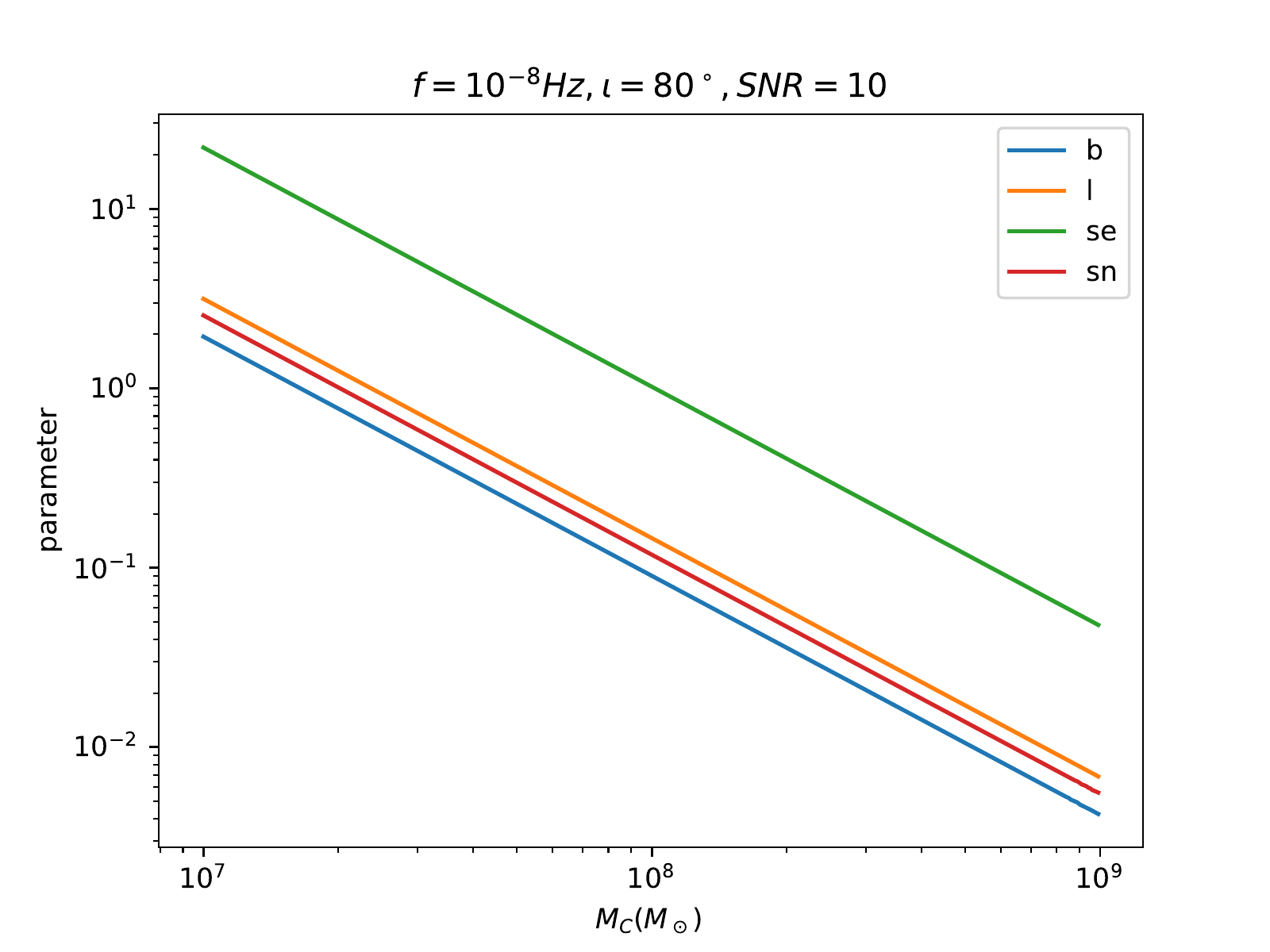}
	\caption{The influence of chirp mass on the capabilities of current PTAs to constrain the non-Einsteinian polarizations. The horizontal axes denote the chirp mass of binaries. The vertical axes denote the constraints of the parameters ($c_b$, $c_{sn}$, $c_{se}$, $c_l$) in equations (\ref{timing_residual}) which represent the capabilities of PTAs to constrain the non-Einsteinian polarizations. The constraints of the parameters are determined by making signal-to-noise ratio (\ref{SNR}) just reach the detection threshold. The distances and celestial positions of the GWs source and the pulsars, as well as the detection threshold remain the same with the previous cases. The constraints are better for the more massive sources.}
	\label{ipta_Mc}
	\end{figure*}

	\emph{The influence of chirp mass $M_c$.}
We consider the chirp mass between $10^7\mathrm{M_{\odot}}$ and $10^9\mathrm{M_{\odot}}$, and the frequency $f = 10\mathrm{nHz}$. The distances and the celestial positions of the binary and the pulsars remain the same with the previous cases. The results are shown in figure (\ref{ipta_Mc}). The constraints are better for the more massive sources. From the equations (\ref{timing_residual}), we can see that the timing residuals depend on the chirp mass by $M_c^{\frac{4}{3}}$. The more massive binaries correspond to the stronger signal for which the constraints are better. As the inclination angle increasing, the lines of $sn$ mode keep stationary, the lines of $b$ and $l$ modes move down, the lines of $se$ mode move upwards. This feature is coincident with the previous cases.

\subsection{Constraints by next generation PTAs}
The eagerly anticipated next generation radio telescope facility, Square Kilometer Array (SKA), will be able to conduct a new era of astronomy, break new ground in astronomical observations. From the simulations based on pulsar population modes, it has been shown that there are up to 14000 normal pulsars and 6000 millisecond pulsars can be discovered by SKA \cite{smits2009pulsar}. The number of pulsars, which can be used in next generation PTAs, will be hundreds, instead of few dozens in current PTAs. In addition to the larger array of pulsars, the timing precision and accuracy will be enhanced remarkably by SKA. With the improved sensitivity of SKA, timing uncertainties of most stable MSPs will be less than 100 ns.

\begin{figure*}[h]
	\centering
	\includegraphics{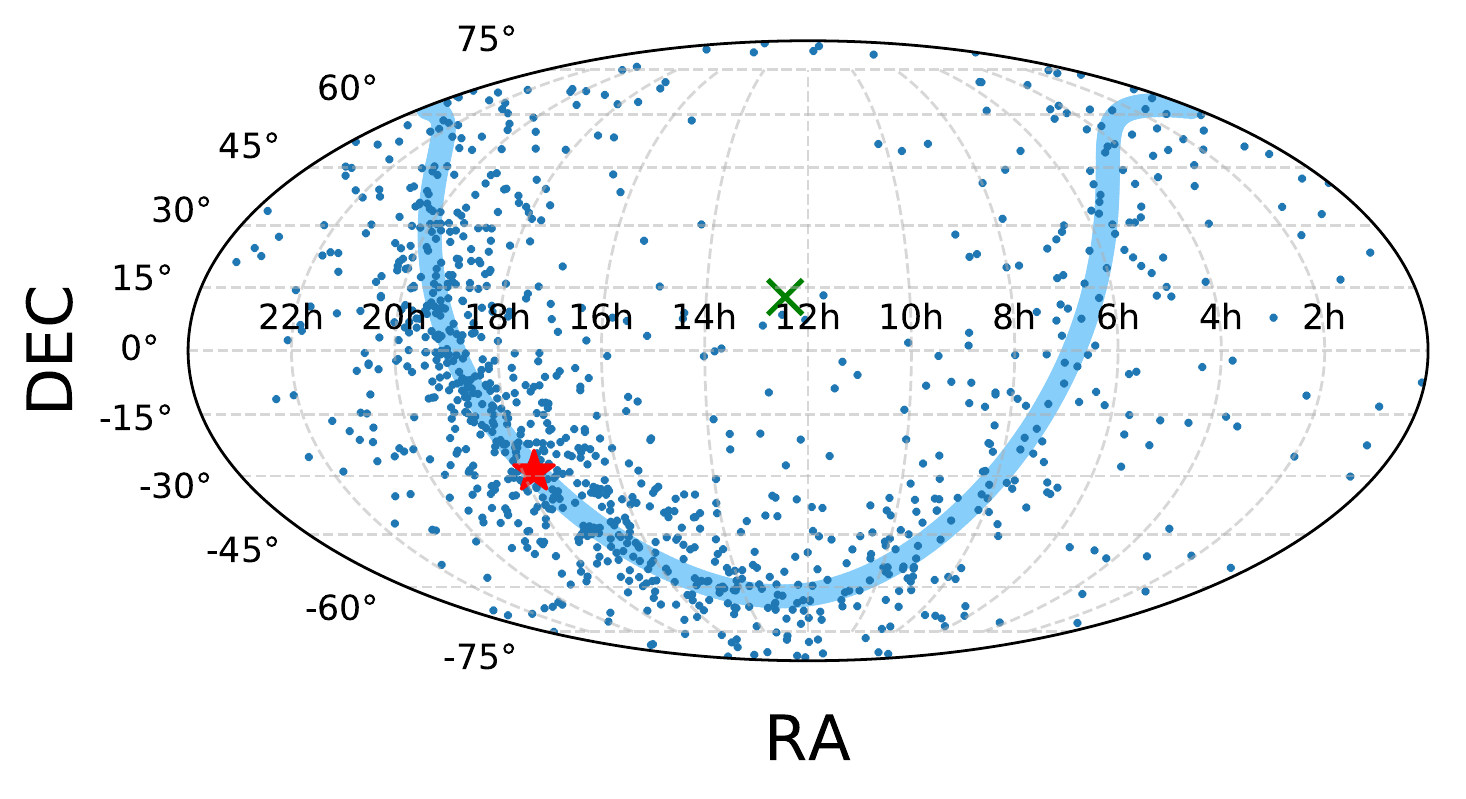}
	\caption{Sky positions of all pulsars in the simulated PTA which is used to represent the next generation PTAs. The PTA considered here is taken from the previous work \cite{wang2017pulsar}. It was constructed using simulated pulsar catalog in \cite{smits2009pulsar} and selecting 1026 MSPs within 3 kpc from us. The dots show the locations of the MSPs constituting the simulated PTA used in this paper. The Milk Way plane is shown as a blue band, and the galactic center is shown as a red star. The green cross denotes the celestial position of gravitational waves source.}
	\label{pulsar_poisiton}
	\end{figure*}

We use the simulated PTA which is taken from the previous work \cite{wang2017pulsar} to represent future PTAs. The PTA considered here was constructed using simulated pulsar catalog in \cite{smits2009pulsar} and selecting 1026 MSPs within 3 kpc from us. Their locations are shown in figure (\ref{pulsar_poisiton}). As in the previous case at which we discussed the current PTAs, we assume an observation with 10-year span and 2-week cadence. The pattern of the observation determines the sensitivity frequency band which is the same with the previous case. An independently and identically distributed zero mean Gaussian noise is considered. The RMS is set to 100 ns for each pulsar \cite{wang2017pulsar}.

\begin{figure*}[h]
	\centering
	\includegraphics[scale=0.5]{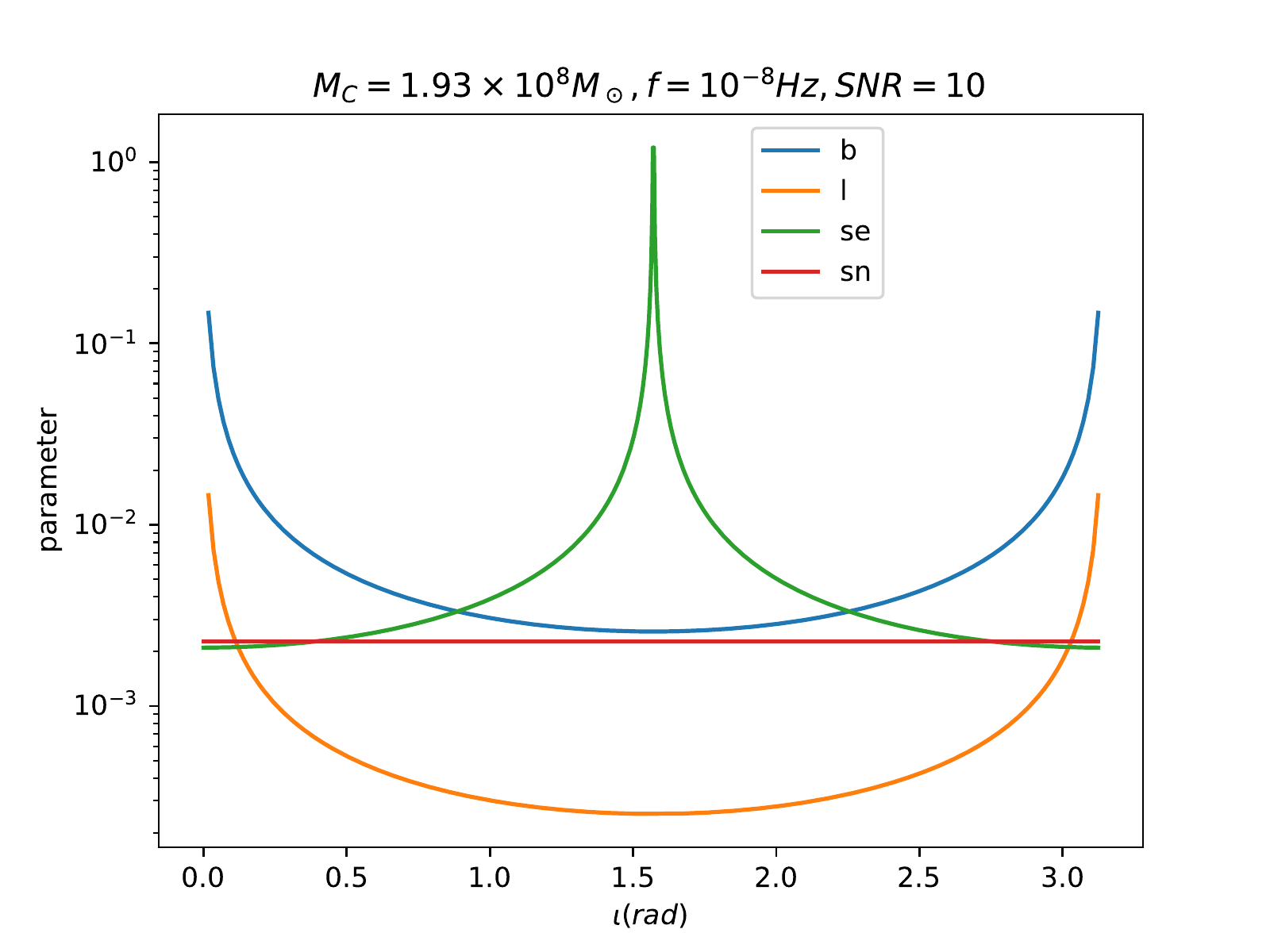}
	\includegraphics[scale=0.5]{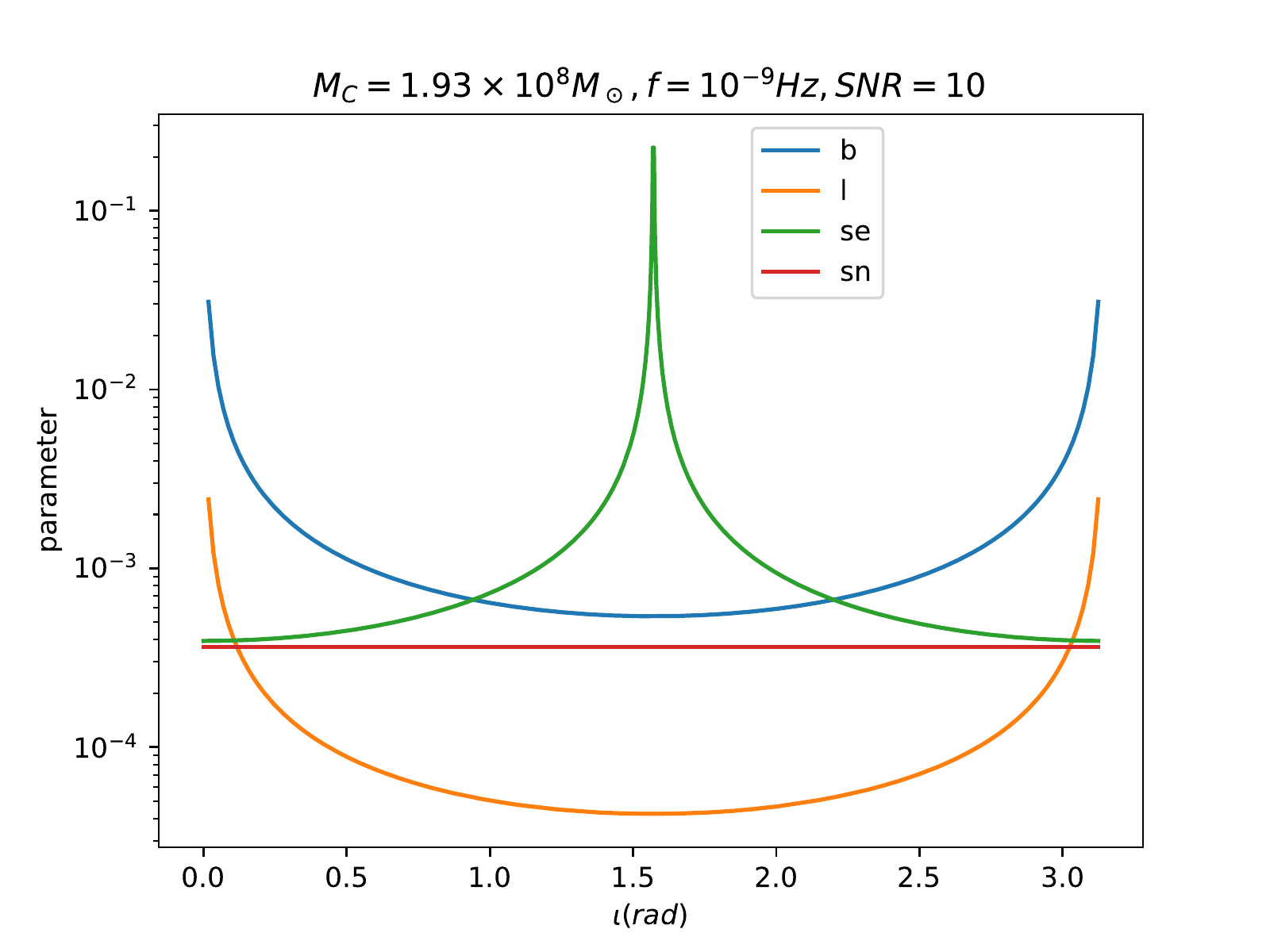} \\
	\includegraphics[scale=0.5]{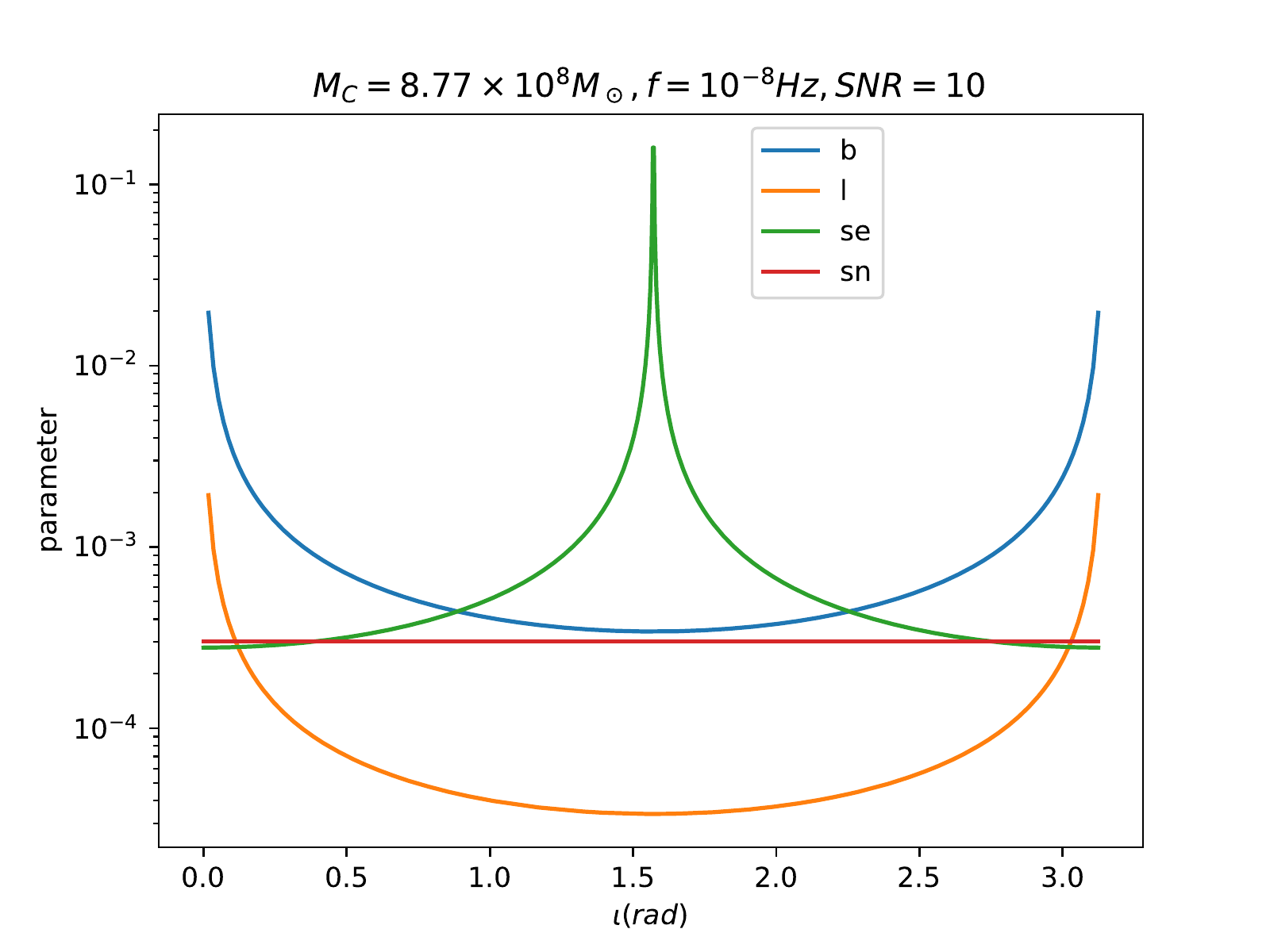}
	\includegraphics[scale=0.5]{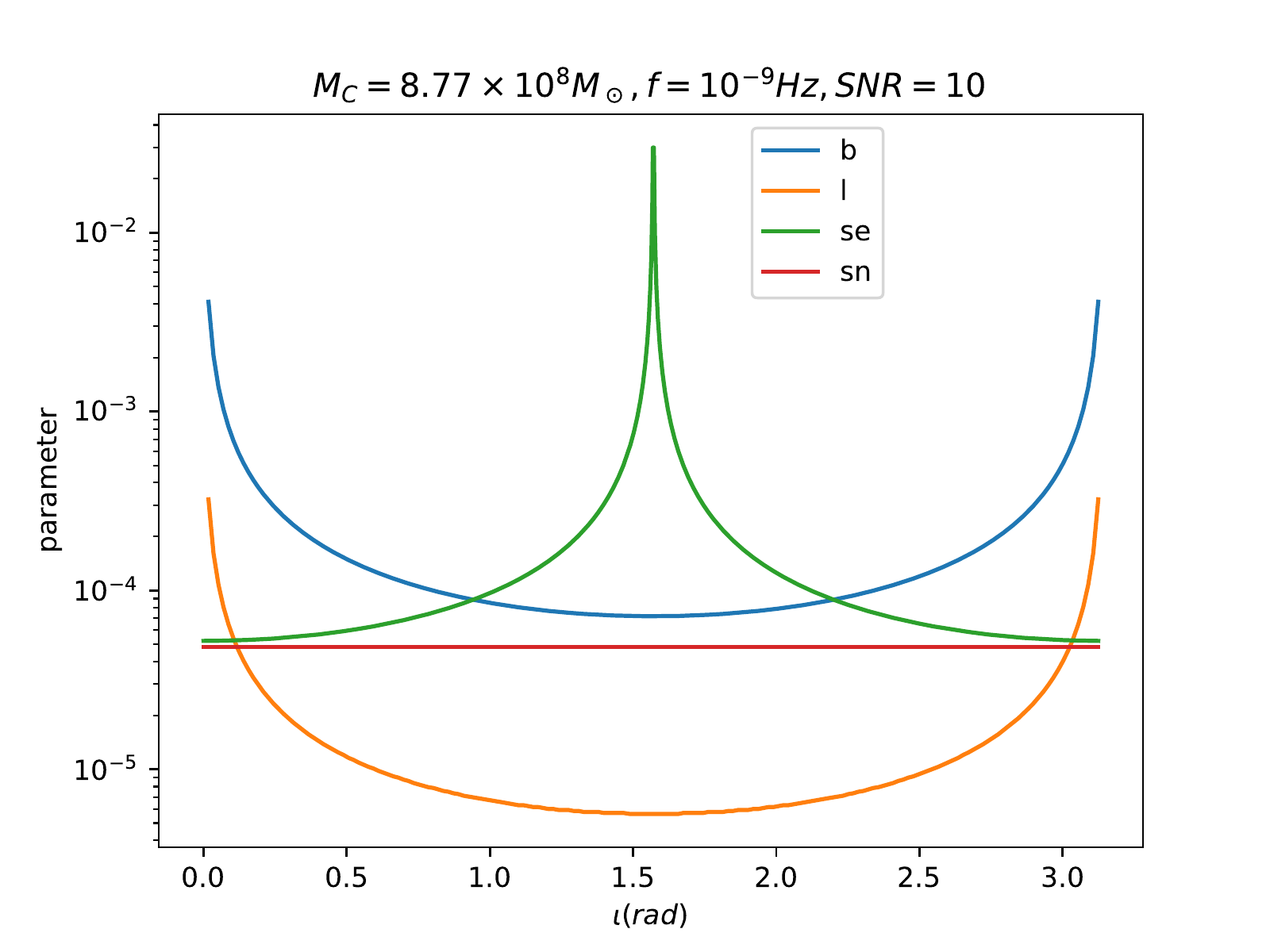}
	\caption{Same with figure (\ref{ipta_i}), but here the current PTA is replaced by the simulated PAT in SKA era.}
	\label{ska_i}
	\end{figure*}

\begin{figure*}[h]
	\centering
	\includegraphics[scale=0.5]{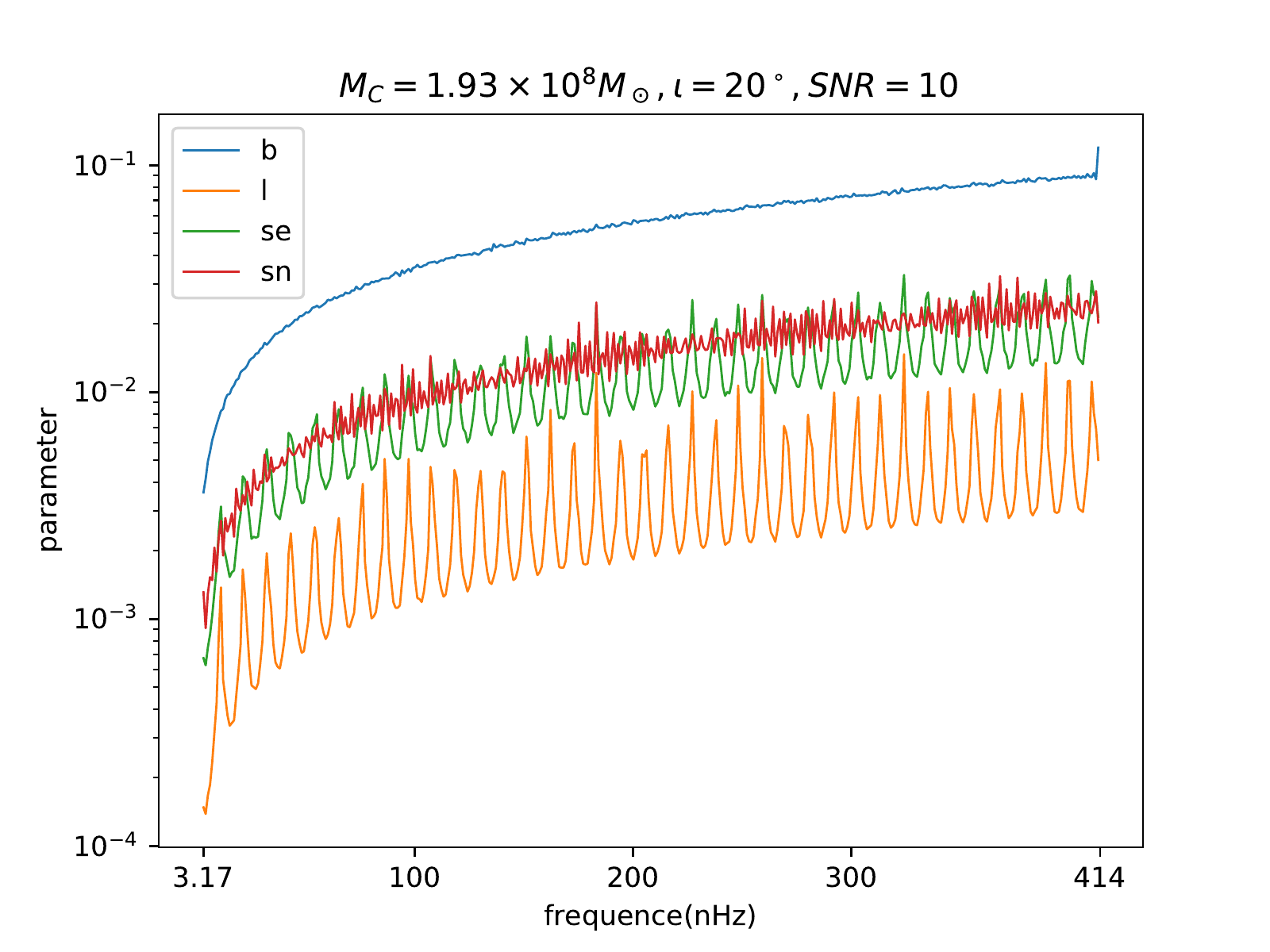}
	\includegraphics[scale=0.5]{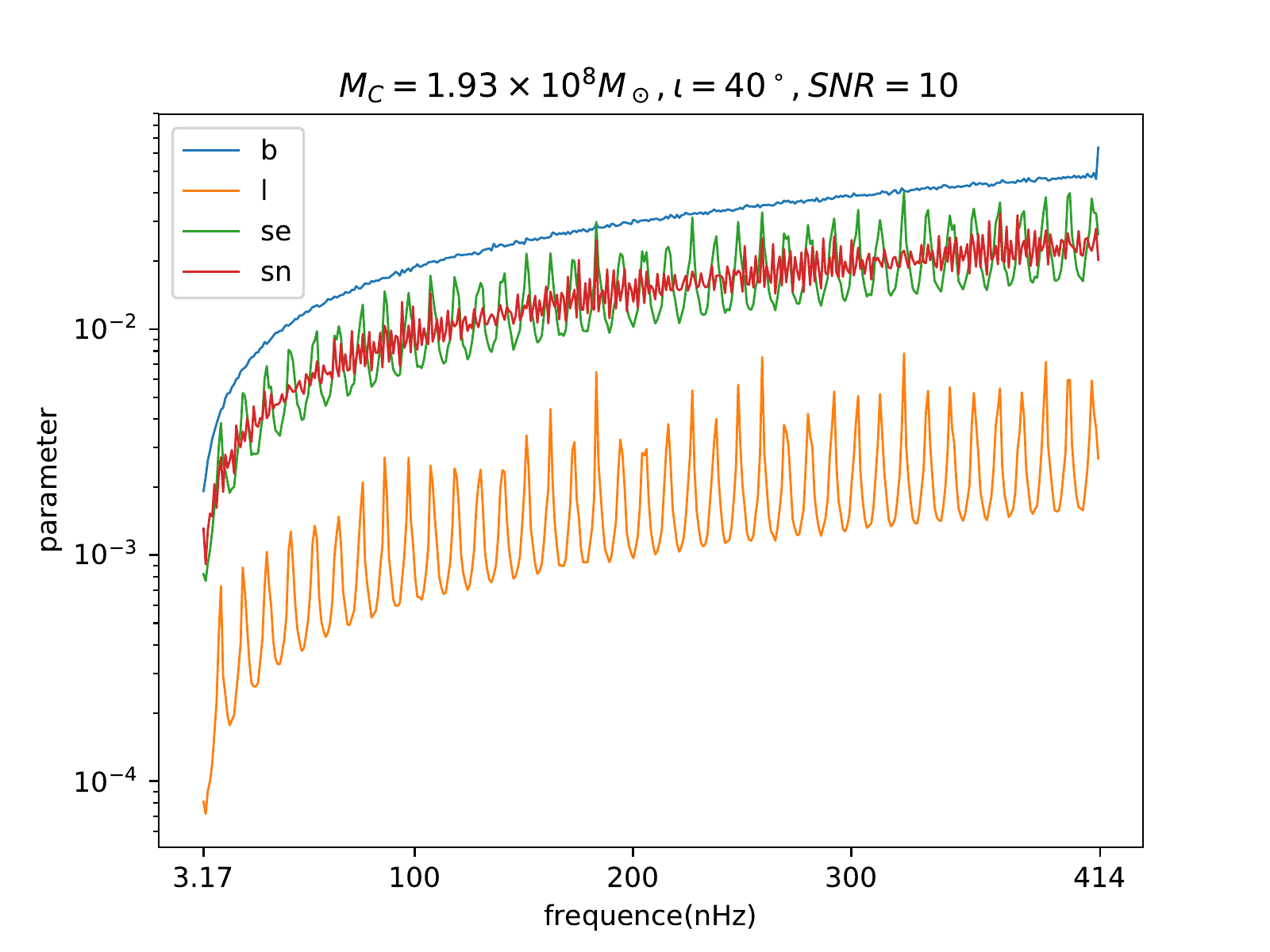} \\
	\includegraphics[scale=0.5]{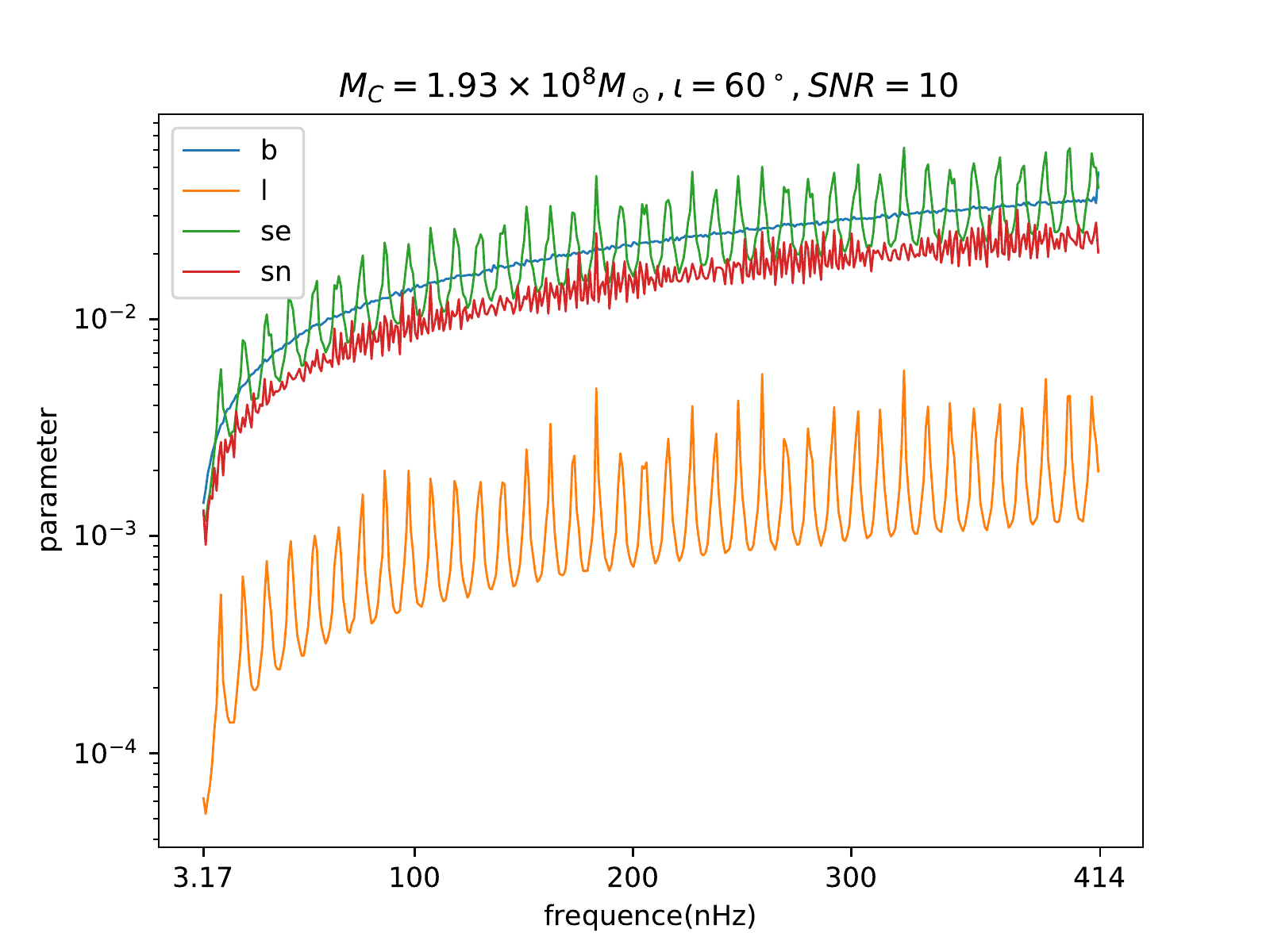}
	\includegraphics[scale=0.5]{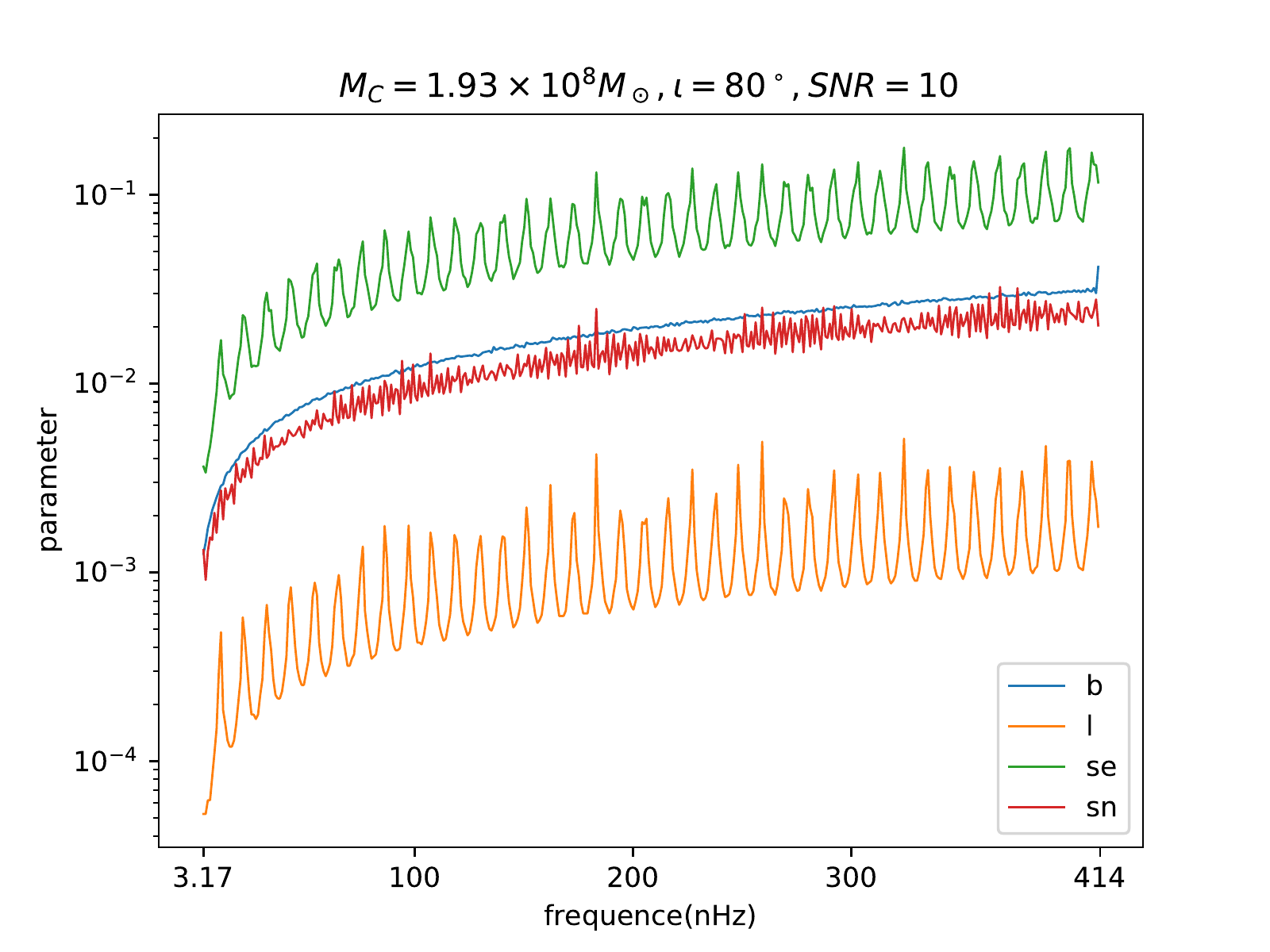}
	\caption{Same with figure (\ref{ipta_f}), but here the current PTA is replaced by the simulated PAT in SKA era.}
	\label{ska_f}
	\end{figure*}

\begin{figure*}[h]
	\centering
	\includegraphics[scale=0.5]{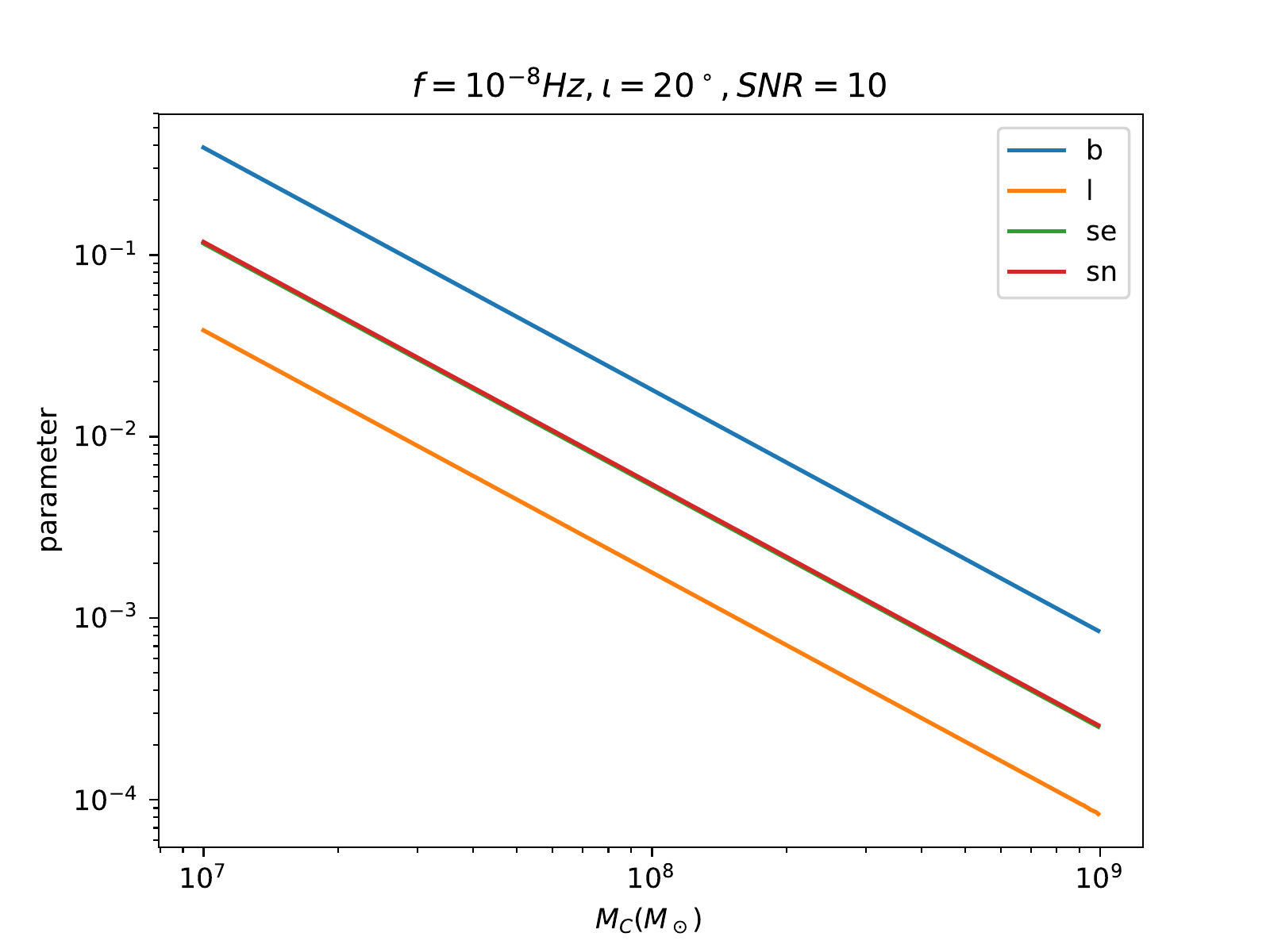}
	\includegraphics[scale=0.5]{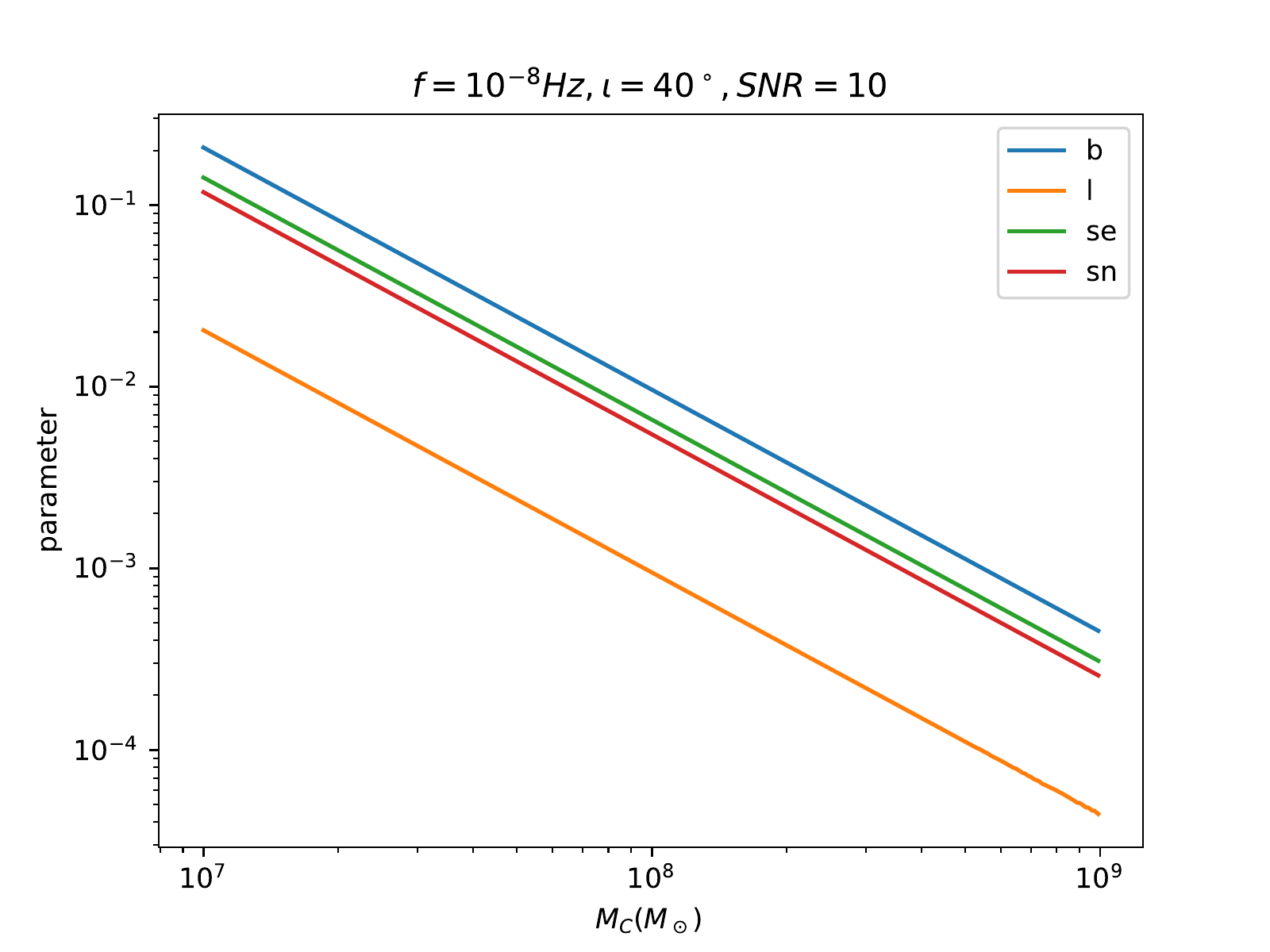} \\
	\includegraphics[scale=0.5]{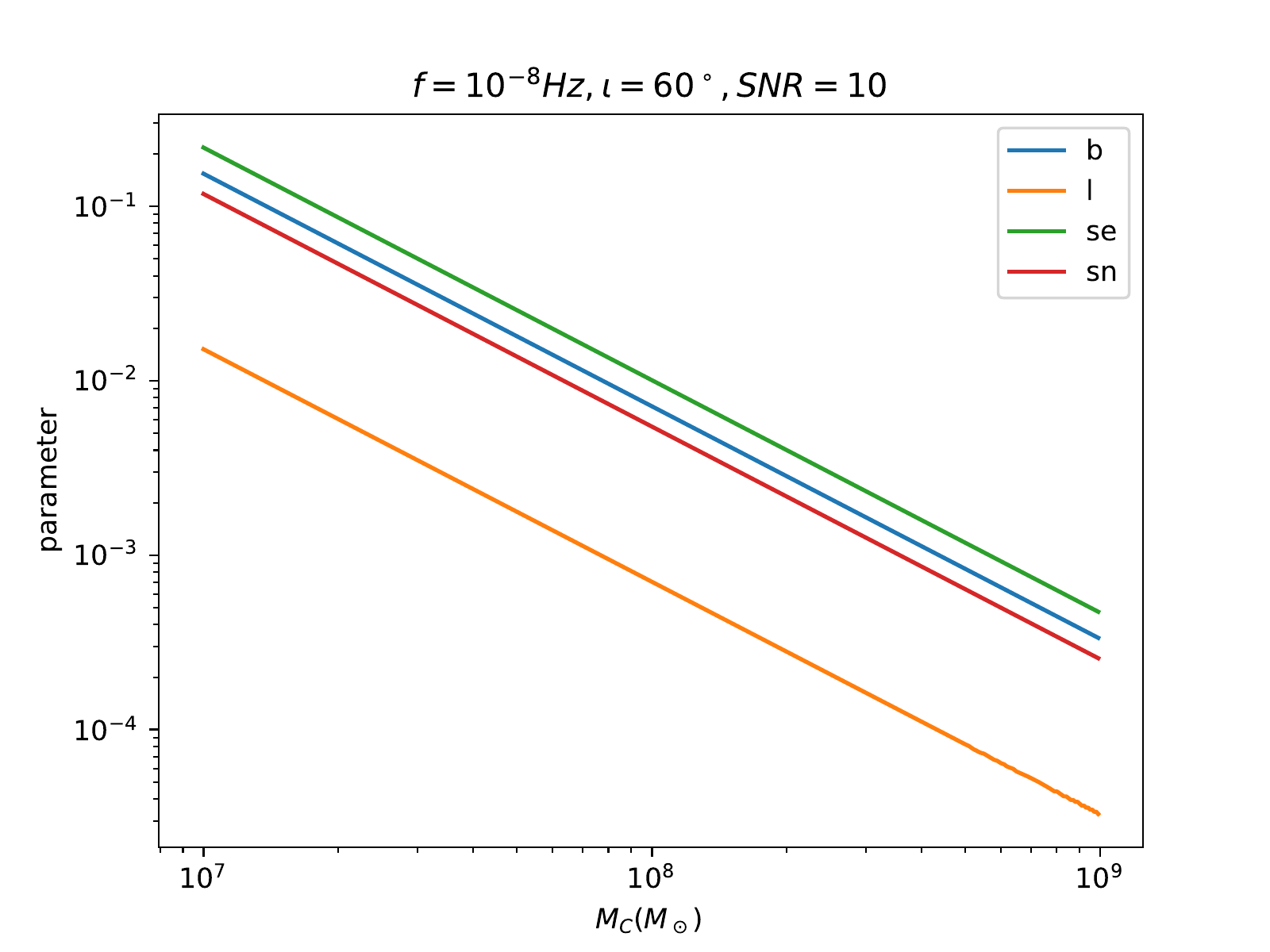}
	\includegraphics[scale=0.5]{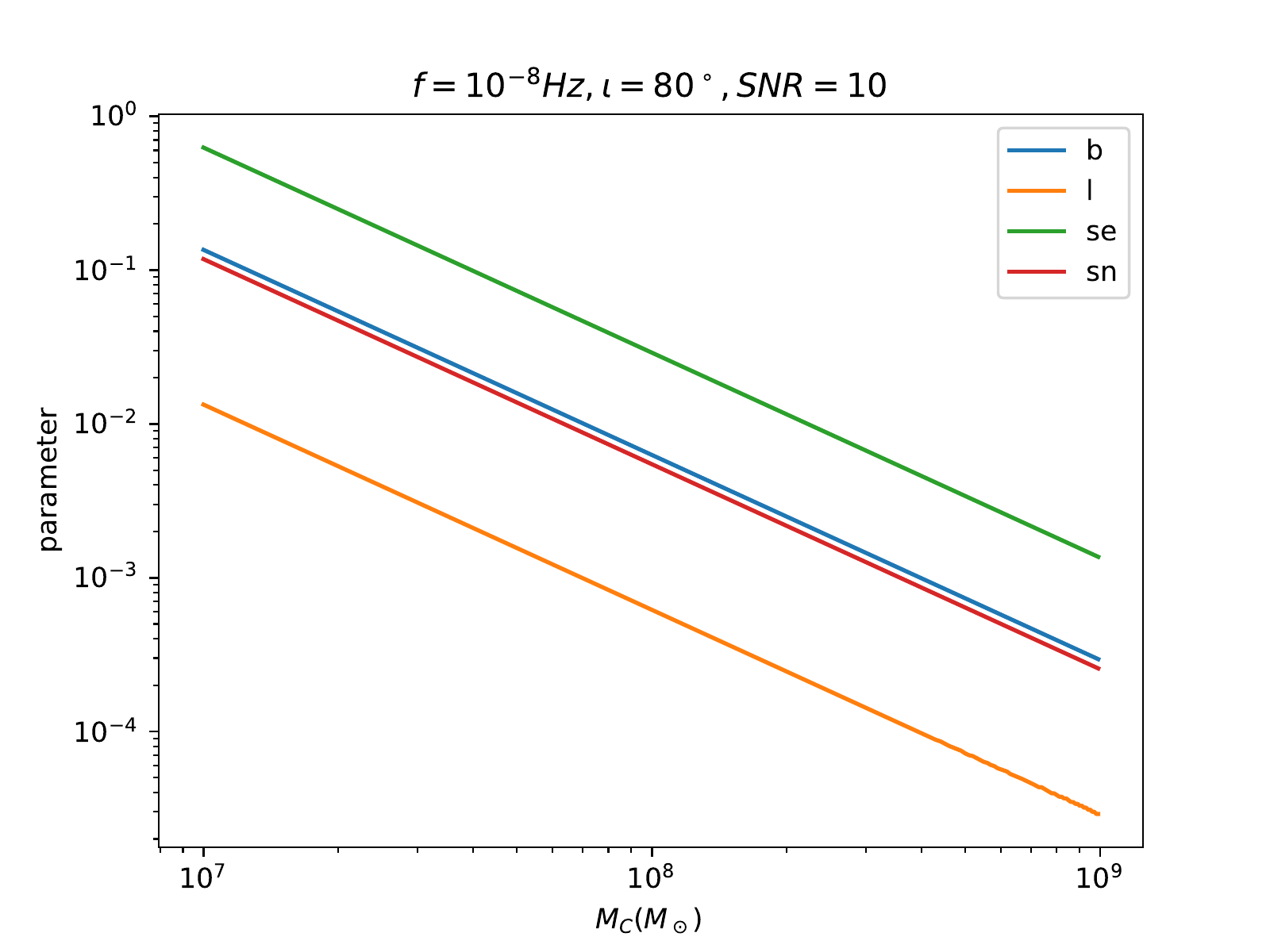}
	\caption{Same with figure (\ref{ipta_Mc}), but here the current PTA is replaced by the simulated PAT in SKA era.}
	\label{ska_Mc}
	\end{figure*}

The analyses are similar with the previous case at which we discussed the current PTAs, except that the pulsars constituting the PTA are substituted and the RMS of noise is better. By the similar analyses, we investigate the influence of chirp mass, orbital frequency and inclination angle on the capabilities of PTAs to constrain the non-Einsteinian polarizations. The capabilities of PTAs to constrain the non-Einsteinian polarizations are represented by the quantities of parameters ($c_b$, $c_{sn}$, $c_{se}$, $c_l$). The position and celestial position of the GWs source, the detection threshold, the pattern of observations which determines the frequency band when we consider the influence of frequency, the range of chirp mass when we consider the influence of chirp mass, as well as the four sets of signals when we consider the influence of inclination angle, are the same with before. The results are shown in figure(\ref{ska_i}), figure(\ref{ska_f}) and figure(\ref{ska_Mc}).

Because of the larger array of pulsars and more accurate measurement, the capabilities of PTAs in SKA era to constrain the parameters in non-Einsteinian polarizations are remarkably enhanced.
Considering the influence of inclination angle, with the optimal inclination angle, for case 1 with $M_{c}=1.93 \times 10^{8} \mathrm{M_{\odot}}$ and $f=10^{-8}\mathrm{Hz}$, the constraints of four non-Einsteinian polarizations are: $c_b = 2.57 \times 10^{-3}$, $c_l = 2.54 \times 10^{-4}$, $c_{se} = 2.10 \times 10^{-3}$, $c_{sn} = 2.27 \times 10^{-3}$. While for case 4 with $M_{c}=8.77 \times 10^{8} \mathrm{M_{\odot}}$ and $f=10^{-9}\mathrm{Hz}$, the constraints of four non-Einsteinian polarizations are: $c_b = 7.17 \times 10^{-5}$, $c_l = 5.63 \times 10^{-6}$, $c_{se} = 5.22 \times 10^{-5}$, $c_{sn} =4.82 \times 10^{-5}$. They are more than one order smaller than those in the corresponding cases with current PTA.

\begin{figure*}[h]
	\centering
	\includegraphics[scale=0.5]{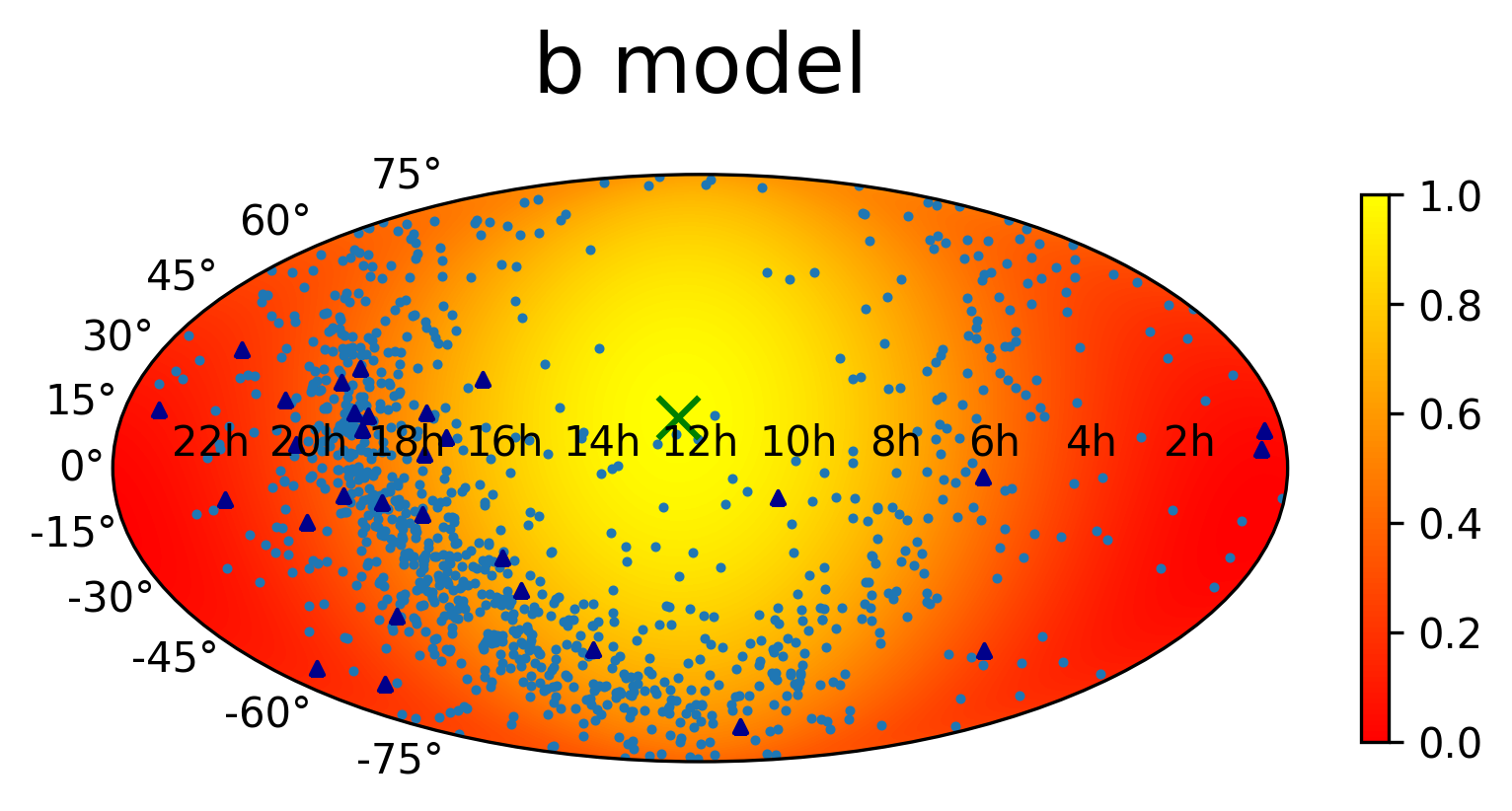}
	\includegraphics[scale=0.5]{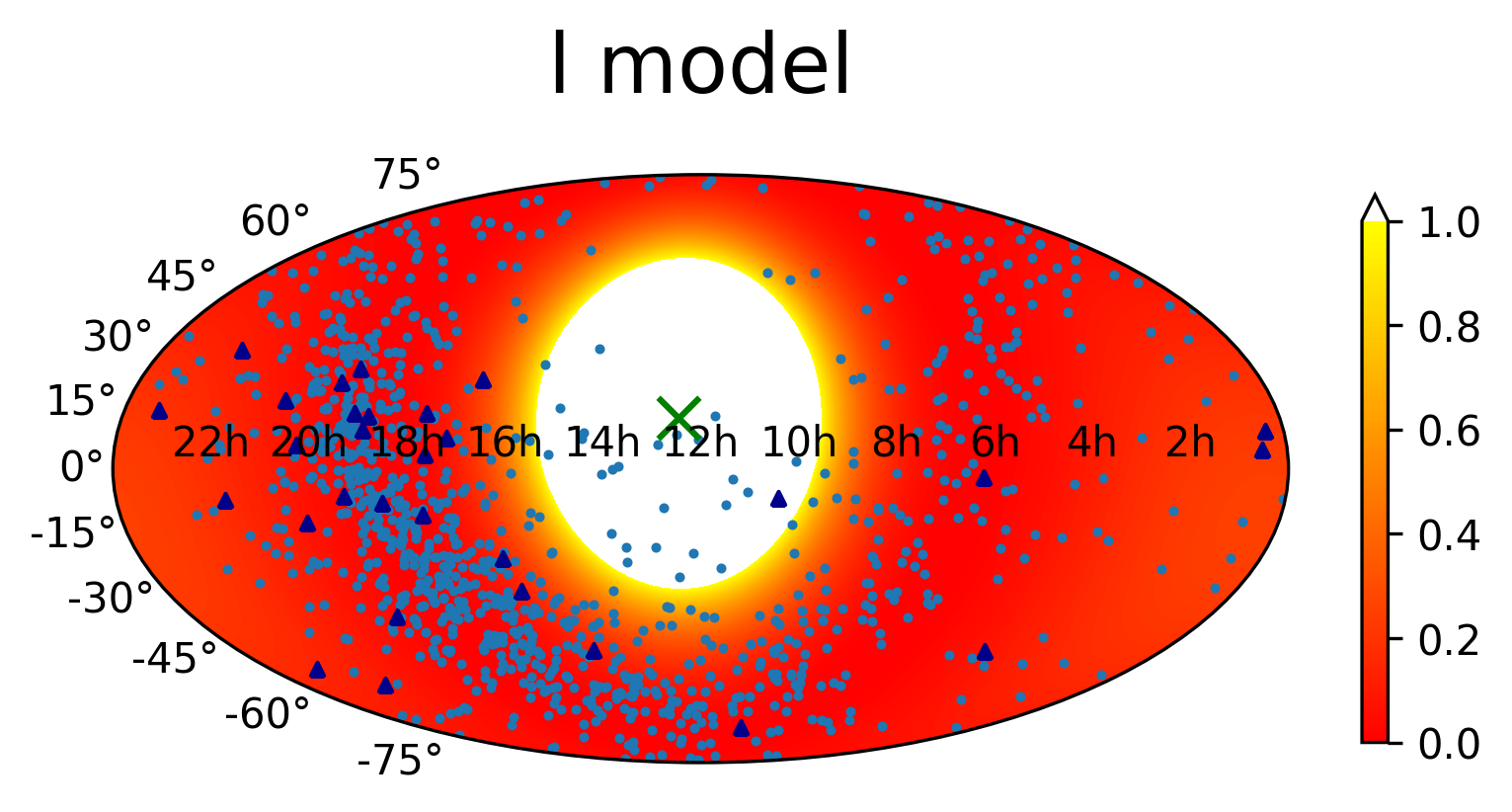} \\
	\includegraphics[scale=0.5]{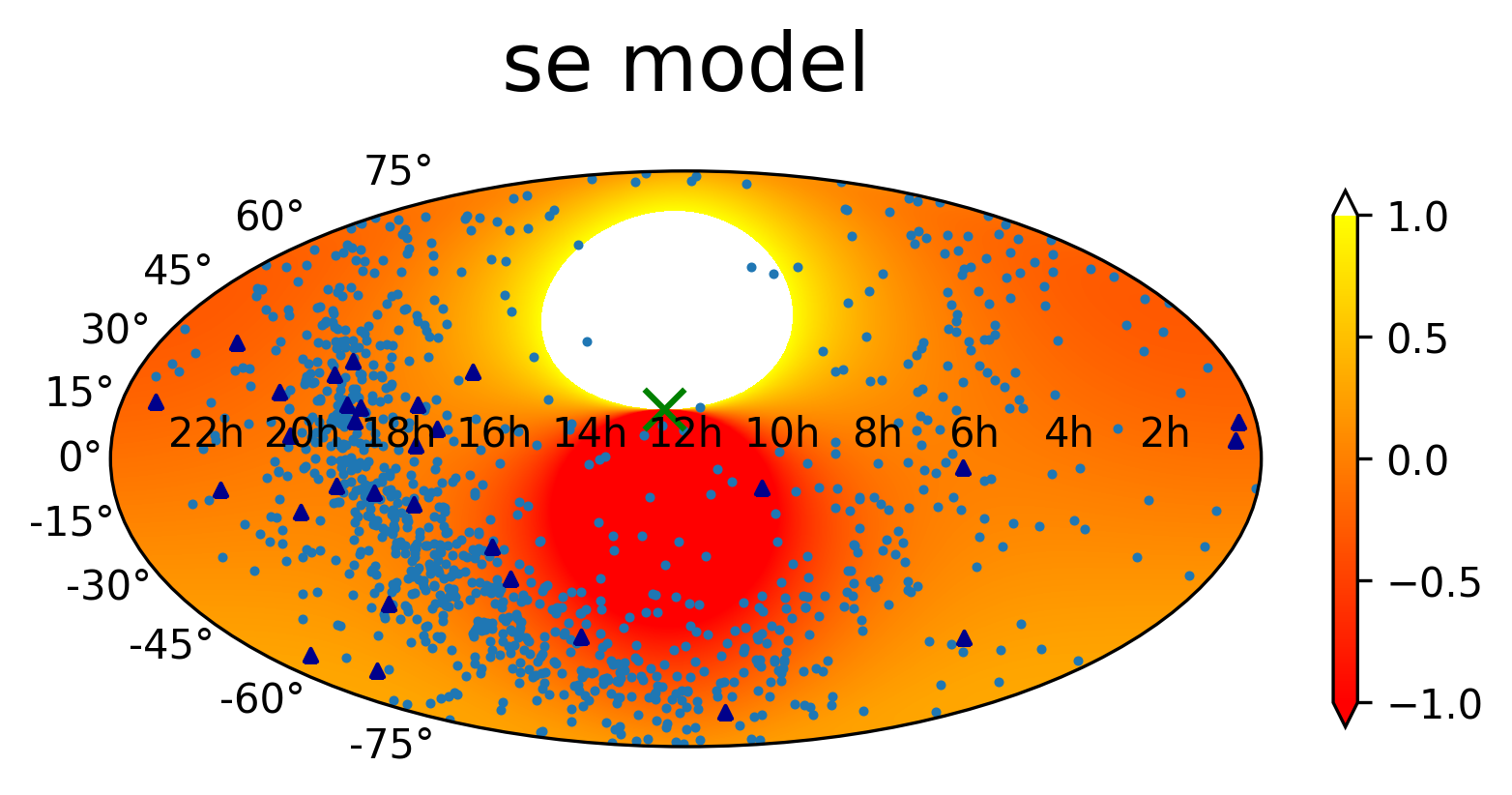}
	\includegraphics[scale=0.5]{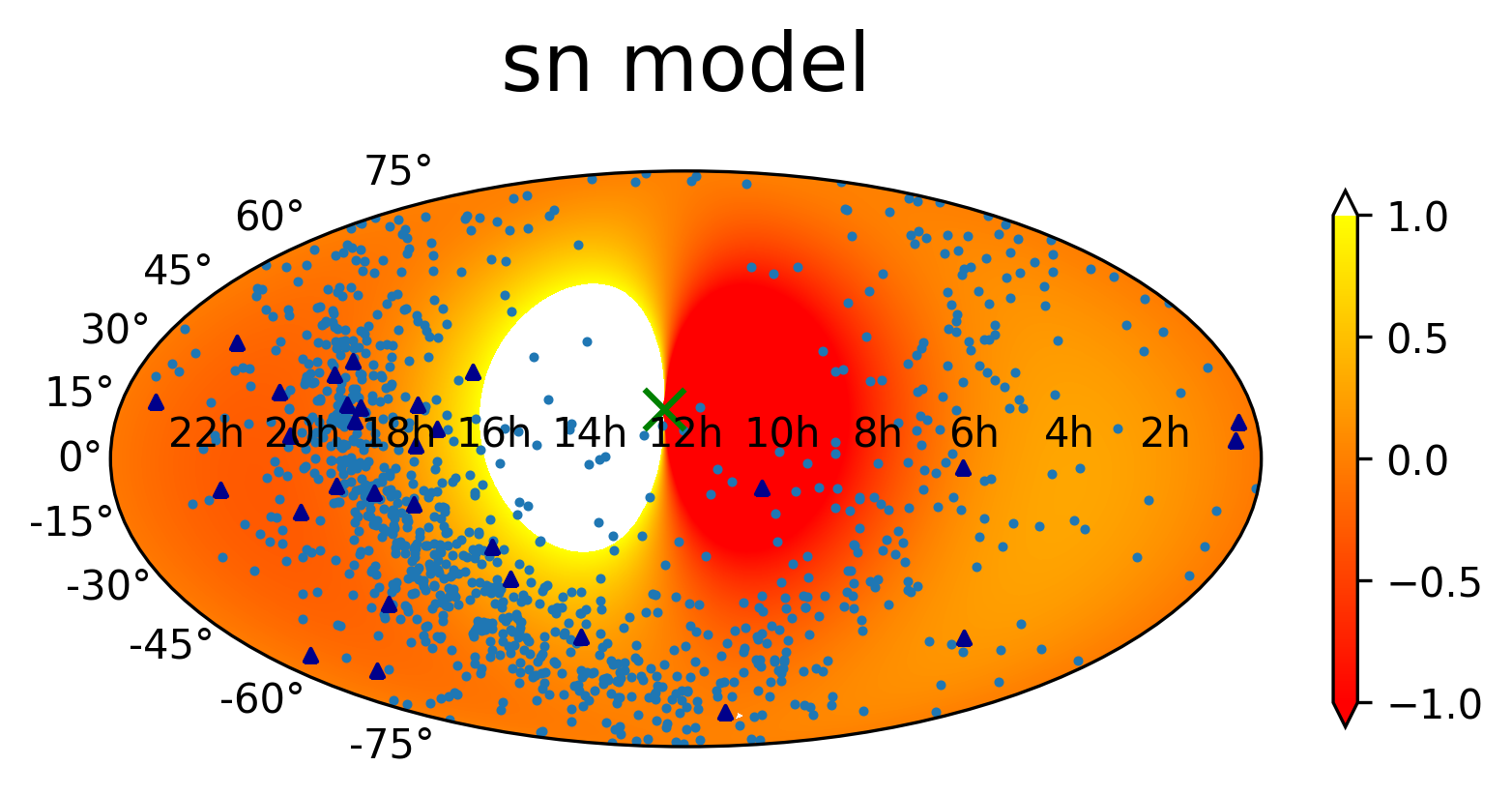}
	\caption{The magnitude of geometrical terms of four modes and celestial positions of pulsars. The magnitude of geometrical terms of all-sky positions can be calculated by equations (\ref{geometry_term_s}), if a specific celestial position of GWs source is given. There is divergence in the geometrical terms of $l$, $se$, $sn$ modes. The red and white holes denote the regions where the values of geometrical terms exceed $1$ or $-1$. The green cross denotes the celestial position of gravitational waves source. The blue dots denote simulated pulsars in SKA era PTA, the dark blue triangles denote pulsars in current PTA. Because of geometrical terms, for a pulsar located at a specific position, the timing residual responses to four polarization modes are different.}
	\label{geometry_term&skyposition}
	\end{figure*}

Comparing the results of the current PTA with the corresponding figures of the future PTA, besides the constraints are better, we can find that the order of four lines has some changes, which means that the order of capabilities to constrain the four modes has changes in the same case. The changes result from the influence of geometrical terms (\ref{geometry_term}). The magnitude of geometrical terms of four modes (for the specific celestial position of the GWs source considered in this work) and celestial positions of pulsars are shown in figure (\ref{geometry_term&skyposition}), in which the blue dots denote simulated pulsars in the future PTA, the dark blue triangles denote pulsars in the current PTA. Because of geometrical terms, for a pulsar located at a specific position, the timing residual responses to four polarization modes are different. The order of capabilities to constrain the four modes will change, if we use different pulsar sets constituting the PTA.

	\begin{table}[H]
	\centering
	\caption{The variance of square of geometrical terms of pulsars. For the specific celestial position of the GWs source, the magnitude of geometrical terms of pulsars can be calculated by equations (\ref{geometry_term_s}). The variance reveals the spread of the values of geometrical terms. The small variance means that the contributions of each pulsar in the PTA to the signal-to-noise ratio are comparable; the large variance means that there are some pulsars which have dominant contributions to the signal-to-noise ratio. }
	\label{var_geoterm}
	\begin{tabular}{lcccc}
		\Xhline{1.2pt}
		& b & l & se & sn \\
		\Xhline{0.6pt}
		Current PTA        & 0.0406 &   0.251            & 0.0912 & 0.621 \\
  		SKA         & 0.0317 & $9.60 \times 10^5$ & 467    & 28.7  \\
		\Xhline{1.22pt}
	\end{tabular}
	\end{table}

There is another noticeable difference. Comparing the figure (\ref{ipta_f}) with the figure (\ref{ska_f}), It is confusing that the features of four modes are quite different in the figure (\ref{ska_f}). We attribute this to the geometrical terms. The timing residual responses to four polarization modes are shown in equations (\ref{timing_residual}). For every pulsar, when we vary the frequency, the changes of amplitude terms in equations (\ref{timing_residual}) are monotonic, but the changes of the part in braces of equations (\ref{timing_residual}) have different cases. Therefore, if there are many pulsars to accumulate in calculations of the signal-to-noise ratio (\ref{SNR}), the fluctuations will be suppressed. However, because of the existence of geometrical terms, it is possible that only a few pulsars in the PTA have dominant contributions to the signal-to-noise ratio (\ref{SNR}). The magnitude of geometrical terms of four modes (for the specific celestial position of the GWs source considered in this work) and celestial positions of pulsars are shown in figure (\ref{geometry_term&skyposition}). With the promotion from the current PTA to the PTA in SKA era, for $b$ polarization mode, the number of pulsars indeed increases, but for other three polarization modes, there is no effective increase of pulsar number. The variance of square of geometrical terms of the pulsars (for the specific celestial position of the GWs source considered in this work) is shown in table (\ref{var_geoterm}). The variance reveals the spread of the values of geometrical terms. The small variance means that the contributions of each pulsar in the PTA to the signal-to-noise ratio are comparable, and the large variance means that there are some pulsars which have dominant contributions to the signal-to-noise ratio. Although each pulsar in the IPTA has comparable contributions to the signal-to-noise ratio, the number of pulsars is not enough to suppress the fluctuations. For the $b$ mode, the variance of the simulated PTA in SKA era is small, the number of pulsars indeed increases, so the fluctuations of $b$ mode are well suppressed. For the $l$, $sn$, $se$ modes, the variance of geometrical terms of pulsars in SKA era PTA is large, which means that, although the array is larger, there are only a few pulsars that have dominant contributions to the signal-to-noise ratio. We can see that the extent of fluctuations is consistent with the quantities of variance for the three modes.

\section{Conclusions}
In metric theories of gravity, GWs can have up to six possible polarizations. Unlike interferometric detectors, a PTA includes a large number of ``arms''. The directional complexity makes PTA superior for detecting different polarization modes of GWs. In this work, we considered the capabilities of current and future PTAs to constrain the non-Einsteinian polarizations. The GW source considered in this paper is a inspiralling supermassive black hole binary in circular orbit. And we neglected the frequency evolution of the binary. We started by considering the responses of timing residuals to the GWs in GR. And then we extended the results to the general polarization modes of GWs. The timing residuals of other four possible polarization modes were shown in the equations (\ref{timing_residual}). The capabilities of PTAs to constrain the non-Einsteinian polarizations are represented by the quantities of parameters ($c_b$, $c_{sn}$, $c_{se}$, $c_l$) in the equations (\ref{timing_residual}) that can make signal-to-noise ratio (\ref{SNR}) just reach the detection threshold. The timing residuals caused by GWs depend on chirp mass, orbital frequency and inclination angle of the binary, as well as distances and celestial positions of the binary and pulsars. We analyzed the influence of chirp mass, orbital frequency and inclination angle on the capabilities of PTAs to constrain the non-Einsteinian polarizations. For the binary, we considered a circular binary system with distance $r = 16.5 \ \mathrm{Mpc}$ and celestial position $(\alpha, \delta) = (3.2594 \ \mathrm{rad}, 0.2219 \ \mathrm{rad})$ located in the Virgo cluster as in the previous work \cite{zhu2016detection}. For the pulsars, we used the same simulated PTA in the previous work \cite{zhu2016detection} to represent current PTAs. The PTA consist of 30 IPTA pulsars as listed in table (\ref{table_IPTA-pulsar}). To represent next generation PTAs, we employed the simulated PTA in the work \cite{wang2017pulsar} which was constructed using simulated pulsar catalog in \cite{smits2009pulsar} and selecting 1026 MSPs within 3 kpc from us. The simulated pulsar catalog \cite{smits2009pulsar} contains the pulsars which come from simulations based on pulsar population modes and can be discovered by SKA.

For the current PTA, we analyzed the influence of the inclination angle on the constraints of the non-Einsteinian modes of gravitational waves, and found that the constraints of $b$, $l$ modes are best when the inclination angle is $90^\circ$ and are divergent when the inclination angle is $0^\circ$ or $180^\circ$, the constraints of $se$ mode are best when the inclination angle is $0^\circ$ or $180^\circ$ and are divergent when the inclination angle is $90^\circ$, the constraints of $sn$ mode have no relations to the inclination angle. With the optimal inclination angel, for case 1 with $M_{c}=1.93 \times 10^{8} \mathrm{M_{\odot}}$ and $f=10^{-8}\mathrm{Hz}$, the constraints of four non-Einsteinian polarizations are: $c_b = 0.0369$, $c_l = 0.0599$, $c_{se} = 0.0737$, $c_{sn} = 0.0492$, and for case 4 $M_{c}=8.77 \times 10^{8} \mathrm{M_{\odot}}$ and $f=10^{-9}\mathrm{Hz}$, the constraints become: $c_b = 0.00106$, $c_l = 0.00217$, $c_{se} = 0.00271$, $c_{sn} = 0.00141$. We also investigated the effects of frequency $f$ of GWs and the chirp mass $M_c$ of binary systems on the constraints of the non-Einsteinian modes, and found that the constraints are better for lower frequency and larger chirp mass.

For the future PTA in SKA era, we found the constraints on the non-Einsteinian modes of gravitational waves are about two orders tighter that those by the current PTA, which is due to larger PTA in SKA era, and the smaller timing uncertainties for each pulsar. For instance, we found that with the optimal inclination angle, the GW source with $M_{c}=1.93 \times 10^{8} \mathrm{M_{\odot}}$ and $f=10^{-8}\mathrm{Hz}$, the constraints of four non-Einsteinian polarizations become: $c_b = 2.57 \times 10^{-3}$, $c_l = 2.54 \times 10^{-4}$, $c_{se} = 2.10 \times 10^{-3}$, $c_{sn} = 2.27 \times 10^{-3}$. While for the GW source with $M_{c}=8.77 \times 10^{8} \mathrm{M_{\odot}}$ and $f=10^{-9}\mathrm{Hz}$ with optimal inclination angle, the constraints are $c_b = 7.17 \times 10^{-5}$, $c_l = 5.63 \times 10^{-6}$, $c_{se} = 5.22 \times 10^{-5}$, $c_{sn} =4.82 \times 10^{-5}$. More interesting, due to effects of the geometrical factors, we found that in SKA era, the constraints on the $l$, $sn$, $se$ modes of GWs are purely dominated by several pulsars.

In the end of this article, we should emphasize that, although we considered only the SMBHB in Virgo Cluster as an example for illustration, the conclusion and the constraints derived in this paper can be simply extended to the SMBHB at any distance. Ignoring the small differences between observed masses and physical masses, and that between observed frequency and physical frequency, the constraints of various polarization modes derived in this paper will be increased by a factor $r/r_{v}$ ($r$ and $r_{v}$ are the luminosity distances of SMBHB and Virgo cluster respectively), since the value of signal-to-noise ratio is inversely proportional to the distance of the GW sources.

\Acknowledgements{acknowledgements:
This work is supported by NSFC No. 11603020, 11633001, 11173021, 11322324, 11653002, 11421303, project of Knowledge Innovation Program of Chinese Academy of Science, the Fundamental Research Funds for the Central Universities and the Strategic Priority Research Program of the Chinese Academy of Sciences Grant No. XDB23010200. }

\InterestConflict{The authors declare that they have no conflict of interest.}





\end{multicols}

\end{document}